\documentclass{aa}  

\usepackage{url}
\usepackage{graphicx}
\graphicspath{{/media/4C2AE0752AE05E08/These/Article/Figures/}}
\usepackage{txfonts}
\usepackage{lscape}
\def\spef{$\mathcal{F}$}
\bibpunct{(}{)}{;}{a}{}{,} % to follow the A&A style (Copied this in)

\begin{document} 

   \title{A glance at the host galaxy of high-redshift quasars using strong damped Lyman-alpha systems as coronagraphs}  
%Characterizing proximate damped Lyman alpha systems that reveal narrow Lyman alpha emission}

   %\subtitle{}

   \author{Hayley Finley \inst{1}
          \and
          Patrick Petitjean \inst{1}
          \and 
          Isabelle P\^{a}ris \inst{2}
          \and
          Pasquier Noterdaeme \inst{1}
          \and
          Jonathan Brinkmann \inst{3}
          \and
          Adam D. Myers \inst{4}
          \and
          Nicholas P. Ross \inst{5}
          \and
          Donald P. Schneider \inst{6, 7}
          \and
		  Dmitry Bizyaev\inst{3}
		  \and
		  Howard Brewington\inst{3}
		  \and
		  Garrett Ebelke\inst{3}
		  \and
		  Elena  Malanushenko\inst{3}
		  \and
		  Viktor Malanushenko\inst{3}
		  \and
		  Daniel Oravetz\inst{3}
		  \and
          Kaike Pan\inst{3}
          \and
          Audrey Simmons\inst{3}
          \and
          Stephanie Snedden\inst{3}
          }

   \institute{Institut d'\,Astrophysique de Paris, CNRS-UPMC, UMR7095, 
              98bis bd Arago, 75014 Paris, France -- \email{finley@iap.fr}
         \and
             Departamento de Astronom\'{i}a, Universidad de Chile, 
             Casilla 36-D, Santiago, Chile
         \and
             Apache Point Observatory, P.O. Box 59, Sunspot, NM 88349-0059, USA
         \and
             Department of Physics and Astronomy, University of Wyoming, Laramie, WY 82071, USA
		 \and 
		 	 Lawrence Berkeley National Laboratory, 1 Cyclotron Road, Berkeley, CA 92420, USA
		 \and
		     Department of Astronomy and Astrophysics, The Pennsylvania State University,
             University Park, PA 16802, USA
         \and
             Institute for Gravitation and the Cosmos, The Pennsylvania State University,
             University Park, PA 16802, USA\\
         %\and
         %	 Other Institutions   \\
             }

   \date{\today}

%\abstract{}{}{}{}{} 
% 5 {} token are mandatory 
\abstract{
We searched quasar spectra from the SDSS-III Baryon Oscillation Spectroscopic Survey (BOSS) for the rare occurrences where a strong damped Lyman-alpha absorber (DLA) blocks the Broad Line Region emission from the quasar and acts as a natural coronagraph to reveal narrow Ly$\alpha$ emission from the host galaxy. We define a statistical sample of 31 DLAs in Data Release 9 (DR9) with log $N(\ion{H}{I}) \geq 21.3\  \text{cm}^{-2}$ located at less than 1\,500~km~$\text{s}^{-1}$ from the quasar redshift. In 25\% (8) of these DLAs, a strong narrow Ly$\alpha$ emission line is observed with flux $\sim$25$\times 10^{-17}\ \rm erg\ s^{-1}\ cm^{-2}$ on average. For DLAs without this feature in their troughs, the average 3-$\sigma$ upper limit is $< 0.8 \times 10^{-17}\ \rm erg\ s^{-1}\ cm^{-2}$. 
Our statistical sample is nearly $2.5$ times larger than the anticipated number of intervening DLAs in DR9 within $1\,500\ \rm km\ s^{-1}$ of the quasar redshift.
%is about half the number in our statistical sample. 
We also define a sample of 26 DLAs from DR9 and DR10 with narrow Ly$\alpha$ emission detected and no limit on the \ion{H}{I} column density to better characterize properties of the host galaxy emission. 
Analyzing the statistical sample, we do not find substantial differences in the kinematics, metals, or reddening for the two populations with and without emission detected. The highly symmetric narrow Ly$\alpha$ emission line profile centered in the HI trough indicates that the emitting region is separate from the absorber. The luminosity of the narrow Ly$\alpha$ emission peaks is intermediate between that of Lyman-alpha emitters and radio galaxies, implying that the Ly$\alpha$ emission is predominantly due to ionizing radiation from the AGN. Galaxies neighboring the quasar host are likely responsible for the majority ($>75$\%) of these DLAs, with only a minority ($<25$\%) arising from \ion{H}{I} clouds located in the AGN host galaxy.
}


   \titlerunning{A glance at the host galaxy of high-z QSOs using strong DLAs as coronagraphs}
   \authorrunning{H. Finley et al.}
   
   \maketitle

%
%________________________________________________________________

\section{Introduction}
Quasar host galaxy properties may provide important constraints on active galactic nucleus (AGN) feedback mechanisms \citep{2009Natur.460..213C, 2012ARA&A..50..455F} and information to better understand the relationship between the growth of the central black hole and the growth of the galaxy \citep[e.g.][]{2000ApJ...539L...9F}. The black hole mass, $M_{\rm BH}$, scales with the stellar mass in the host bulge, $M_{\rm bulge}$, and the stellar velocity dispersion, $\sigma$, linking galaxies' star formation history to the evolution of their supermassive black holes \citep{2000ApJ...539L..13G}. The bulk of star formation in the bulge may occur simultaneously with the maximum black hole growth \citep{2008ApJ...684..853L}, since both processes rely on reservoirs of gas brought to the center by gas-rich mergers and disk instabilities. The growth phase ends when no gas remains, if, for example, star formation consumes all the gas. Alternatively, the AGN may blow the gas outside its host galaxy \citep{2006MNRAS.373L..16F}, curtailing star formation. In this case, the AGN must have sufficient luminosity for the radiation-pressure force outwards to exceed the gravitational force inwards. Untangling the black hole growth and star formation processes is crucial for explaining how galaxies arrive at their present-epoch characteristics. 

For nearby quasar host galaxies, it is possible to obtain detailed observations. Using an integral-field spectrograph, \citet{2012MNRAS.426.1574D} traced the kinematics of ionized gas and identified an AGN-driven outflow disrupting the star-forming gas in NGC 1266 (D = 30 Mpc). Similar studies of nearby quasars are abundant enough to permit a statistical analysis of the $M_{\rm BH}$~--~$M_{\rm bulge}$, $M_{\rm BH}$~--~$\sigma$, and $M_{\rm BH}$~--~$L$ scaling relations in samples compiled from the literature \citep{2009ApJ...698..198G, 2011MNRAS.412.2211G, 2013ApJ...764..184M}. 
The challenge of resolving quasar host galaxies can be overcome at $\text{z} < 0.5$ with deep ground-based imaging (\citealt{2013arXiv1302.1366K}; see also \citealt{2001A&A...371...97M}), at $\text{z} \approx 1$ and $\text{z} \approx 2$ with Hubble Space Telescope imaging \citep{2013MNRAS.429....2F}, and at $z \approx 2 - 3$ with adaptive optics \citep{2008ApJ...673..694F, 2013arXiv1301.1129W}. 
%\textbf{By targeting gravitationally lensed quasars, \citet{2006ApJ...640..114P, 2006ApJ...649..616P} were able to determine host galaxy structures and measure $M_{\rm BH} / M_{\rm bulge}$ ratios at $z > 2$.}
Targeting gravitationally lensed quasars is a particularly effective method for obtaining host galaxy properties at high redshift, up to $z = 4.5$ \citet{2006ApJ...640..114P, 2006ApJ...649..616P}.
These observations at different epochs provide essential constraints on the evolution of quasar host galaxies. 

%By taking advantage of gravitational lensing and using lensing models to disentangle the bright central AGN from the extended host, Peng et al were able to determine the host galaxy structure and measure the $M_{\rm BH} / M_{\rm bulge}$ ratio for a sample of $z > 2$ quasars
%By taking advantage of gravitational lensing, 
%to use lensing models to disentangle the bright central AGN from the extended host,

Millimeter astronomy has been essential for probing quasar host galaxies at higher redshifts. The detection of CO emission in the z = 4.7 quasar BR120-0725 \citep{1996Natur.382..428O}, along with continuum emission due to dust, confirmed that high-redshift quasars are associated with galaxies. This same quasar host galaxy also shows strong Ly$\alpha$ emission \citep{1996Natur.380..411P}, indicating enhanced star formation or elevated ionization from the quasar flux. Recent observations with the Atacama Large Millimeter/submillimeter Array \citep[ALMA;][]{2009IEEEP..97.1463W} of [\ion{C}{II}] line emission in the BR120-0725 host galaxy suggest an outflow of gas ionized by the quasar \citep{2012ApJ...752L..30W}.
At this early epoch, quasar hosts have both vigorous star formation and central black holes that accrete at the Eddington limit \citep{2013arXiv1302.4154W}. Comparing the star formation rate to the black hole accretion rate indicates which process is dominant. For example, \citet{2013arXiv1302.1587W} found that the $z \sim 6.4$ quasar host galaxy J2329-0301 has an uncharacteristically low ratio of star formation to black hole accretion, implying that the quasar has shut off star formation in the galaxy, potentially by photo-ionizing the diffuse gas. Indeed, this quasar has extended Ly$\alpha$ emission across at least 15 kpc \citep{2009MNRAS.400..843G, 2011AJ....142..186W}.

At the $z = 2 - 3$ epoch, when the quasar luminosity function peaks \citep[e.g.,][]{2006AJ....131.2766R}, spectroscopy remains the best tool for investigating host galaxies. An ultra-deep, blind, spectroscopic long-slit survey revealed a young stellar population in the host galaxy of a $z=3.045$ quasar, the formation of which coincides with the quasar activity \citep{2013arXiv1302.2623R}. With spectroscopy, it is also possible to study absorption systems, such as associated narrow absorption lines (NALs), that probe the gaseous environments close to quasars. Associated NAL systems at $z \sim 2.4$ tend to have solar or super-solar metallicities and potentially arise from AGN-driven outflows of gas from the interstellar medium \citep[ISM;][]{1994A&A...291...29P, 2004MNRAS.351..976D}. However, it is difficult to differentiate between gas associated with the AGN, gas from the ISM of the host galaxy, and gas located in galaxies clustered around the AGN host galaxy that is revealed by proximate damped Lyman-alpha absorption systems (DLAs). Sixteen such systems at similarly high redshifts and within $\Delta v < 3\,000\ \text{km s} ^{-1}$ of 
their background quasar have sub-solar metallicities \citep{2010MNRAS.406.1435E}. 

In a newly exploited technique for accessing the quasar host galaxy, DLAs at the redshift of the quasar, hereafter called associated DLAs, can act as natural coronagraphs, completely absorbing the broad Ly$\alpha$ emission from the central AGN. In the absence of broad Ly$\alpha$ emission, it is possible to observe a narrow Ly$\alpha$ emission line from a source at approximately the same redshift as the quasar in the DLA trough. For this situation to occur, the absorber must cover the $\sim$1~pc Broad Line Region (BLR) without also obscuring extended Ly$\alpha$ emission, which can be a result of both UV flux from the AGN that ionizes surrounding gas in the Narrow Line Region (NLR) and star formation activity within the host galaxy.
%This Ly$\alpha$ emission can be the result of both UV flux from the AGN that ionizes surrounding gas in the Narrow Line Region (NLR) and star formation activity within the host galaxy. 
% This Ly$\alpha$ emission can be due both to UV flux from the AGN that ionizes surrounding gas and provokes Ly$\alpha$ emission in the Narrow Line Region (NLR) and to star formation activity within the host galaxy. 
A striking individual detection of this phenomenon is presented in \citet{2009ApJ...693L..49H}; the four other previous detections of narrow Ly$\alpha$ emission in the trough of a DLA near the quasar redshift all have lower Ly$\alpha$ luminosities \citep{1993A&A...270...43M, 1998A&A...330...19M, 1995qal..conf...55P, 2002A&A...383...91E}. 

The serendipitous alignment of a DLA absorber along the quasar line-of-sight near enough to the host galaxy to reveal narrow Ly$\alpha$ emission in the DLA trough is a rare occurrence. \citet{2011MNRAS.412..448E} exploited the statistical power of the Sloan Digital Sky Survey \citep[SDSS;][]{2000AJ....120.1579Y} to investigate the characteristics of $\Delta v < 10\,000\ \text{km s}^{-1}$ proximate DLA absorbers in a sample compiled from Data Release 5 \citep{2007ApJS..172..634A}. Only one DLA in the sample revealed narrow Ly$\alpha$ emission \citep[reported in][]{2009ApJ...693L..49H}, and no emission signal was seen in the stacked spectrum of 29 $\Delta v \leq 3\,000 \text{ km s}^{-1}$ DLAs. 

Thanks to the increased sample size of quasar spectra in the SDSS-III \citep{2011AJ....142...72E} Baryon Oscillation Spectroscopic Survey (BOSS, \citealp{2013AJ....145...10D}), several associated DLAs with unprecedentedly strong narrow Ly$\alpha$ emission superimposed on their troughs were discovered. In this article, we construct a statistical sample of associated DLAs from the SDSS Data Release 9 \citep[DR9; ][]{2012ApJS..203...21A}. We study the emission and absorption properties of systems in our sample with the goal of characterizing any differences between the populations with and without emission detected in the BOSS spectra. We also characterize the emission properties of a sample of associated DLAs that reveal narrow Ly$\alpha$ emission identified thus far in the survey by visual inspection.

Section 2 describes the definition of a complete statistical sample. We then calculate how many intervening DLAs we expect to find at the quasar redshift and consider the possibility of an overdensity (Section 3). Sections 4 and 5 respectively characterize the two DLA populations and the emission properties. In Section 6, we present several associated DLAs with a covering factor unequal to one and consider the implications. We discuss the results in the context of our two-population scenario in Section 7.

We use a $\Lambda$CDM cosmology with $\Omega_{\Lambda} = 0.73$, $\Omega_{\text{m}} = 0.27$, and $\rm H_{0} = 70\ km\ s^{-1}\ Mpc^{-1}$ \citep{2011ApJS..192...18K}.

%
%__________________________________________________________________

\section{Sample Definition}

\begin{figure*}
\centering
\includegraphics[width=18cm]{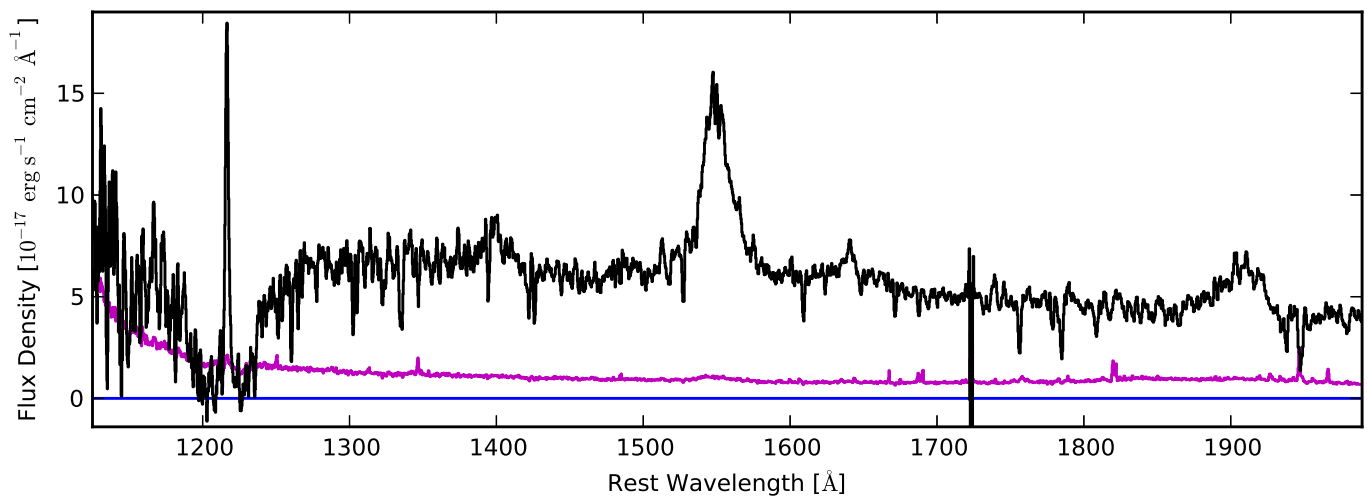}
\caption{Example of a quasar spectrum (SDSS J1256+3506) with an associated DLA that reveals narrow Ly$\alpha$ emission. The 1-$\sigma$ error on the spectral flux density is shown in magenta. }
\label{emExample}
\end{figure*}

\subsection{Strong associated DLAs}

The quasar lines-of-sight were observed as part of SDSS-III BOSS. From 2009 -- 2014, this survey will take spectra of over 150\,000 quasars at $\langle z \rangle \sim 2.5$ and 1.5 million galaxies at $\langle z \rangle \sim 0.7$ with the primary goal of detecting baryon acoustic oscillations at different cosmological times. A dedicated 2.5~m telescope \citep{2006AJ....131.2332G} conducts the imaging and spectroscopy. SDSS-III uses the same camera \citep{1998AJ....116.3040G} as its predecessor surveys \citep{2000AJ....120.1579Y} to obtain images in five broad bands, $ugriz$ \citep{1996AJ....111.1748F}. Twin multi-object fiber spectrographs designed for BOSS collectively span $\sim$3\,600 -- 10\,000~\AA\ with cameras individually optimized for observing at blue and red wavelengths \citep{2012arXiv1208.2233S}. Their resolving power varies from $\sim$1500 -- 3000 across the wavelength range, such that the instrument velocity dispersion is consistently $\sim$150~$\text{km s}^{-1}$. \citet{2012AJ....144..144B} describe the spectroscopic data reduction pipeline.

The Data Release 9 Quasar (DR9Q) catalog \citep{2012A&A...548A..66P} includes 87\,822 quasars detected over 3\,275 $\text{deg}^{2}$, 61\,931 of which are at $z > 2.15$. A combination of target selection methods \citep{2012ApJS..199....3R} achieved this high surface density of high-redshift quasars. The spectra are publicly available as part of DR9. \citet[][ hereafter N12]{2012A&A...547L...1N} provide a catalog of DLAs and sub-DLAs with log$\ N(\ion{H}{i}) \geq 20.0$ detected from an automatic search along the DR9Q lines-of-sight. We use an extended version of the N12 catalog that includes automatic detections up to $+3\,000 \text{ km s}^{-1}$ beyond the quasar redshift. 

We select strong associated DLAs so that the Ly$\alpha$ trough is large enough to completely mask the broad Ly$\alpha$ emission line from the quasar (Figure \ref{emExample}). To gather candidates for a statistical sample of these strong associated DLAs, we search the extended N12 catalog for DLAs that are within $\pm 3\,000 \text{ km s}^{-1}$ of the quasar, have log$\ N(\ion{H}{i}) \geq 21.2$, and arise in lines-of-sight with balnicity index zero and BAL\_FLAG\_VI$=$0 (see \citet{1991ApJ...373...23W} for a discussion of the balnicity index). We exclude quasar spectra with Broad Absorption Lines (BALs), since they could blend with the DLA and contaminate our \ion{H}{I} column density measurements. N12 uses quasar redshifts assigned during the visual inspection, which have an uncertainty of about $500 \text{ km s}^{-1}$ \citep{2012A&A...548A..66P}. Due to this uncertainty on the quasar redshift, we allow the velocity difference limit to extend redward of the quasar, so as not to miss any associated DLAs by stopping the automatic search at the quasar redshift. The column density lower limit was chosen because the zero-level in the trough of the Voigt profile fit spans $1\,000 \text{ km s}^{-1}$ for log$\ N(\ion{H}{i}) \geq 21.2$. Extending to lower column densities would obscure our ability to clearly detect Ly$\alpha$ emission from the host galaxy in the trough. Applying these criteria returns 41 DLAs.

When an associated DLA reveals strong narrow Ly$\alpha$ emission, the N12 automatic search may wrongly identify two consecutive DLAs. To recover these split DLAs, the extended N12 catalog is searched for lines-of-sight that have more than one DLA detected within $\pm 3\,000 \text{ km s}^{-1}$ of the quasar, balnicity index zero and BAL\_FLAG\_VI$=$0. No limit is placed on the column density. Potential split DLAs must be confirmed by visually examining the associated metal absorption lines. Genuine split DLAs have one set of absorption lines that approximately align with the narrow Ly$\alpha$ emission peak, whereas two sets of metal lines are present for consecutive DLAs. This procedure finds 9 associated DLAs split by narrow Ly$\alpha$ emission, bringing the total number of log$\ N(\ion{H}{i}) \geq 21.2$ DLAs within $\pm 3\,000 \text{ km s}^{-1}$ of the quasar to 50.

\subsection{Measuring DLA column densities and emissions}

\begin{figure}
\centering
%\resizebox{\hsize}{!}{\includegraphics{combSpec_csNoEmNo22}}
\includegraphics{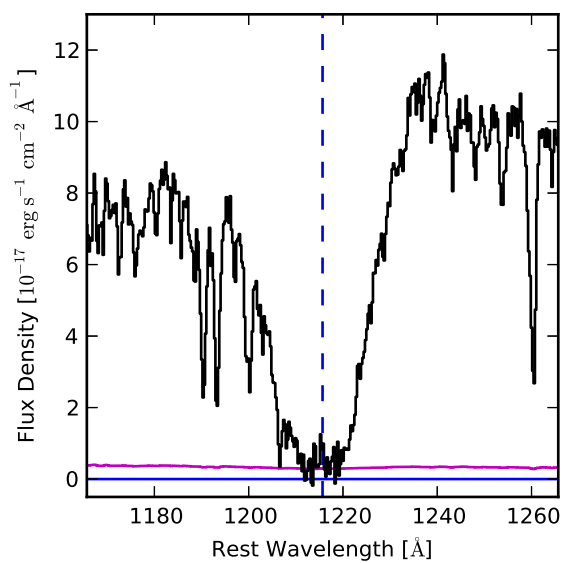}
\caption{Combined spectrum for the twenty-three DLAs in the statistical sample with no Ly$\alpha$ emission detected at the 4-$\sigma$ level. The spectra are combined with an inverse variance weighted average. The zero-flux density level in the DLA trough is approximately at the level of the error on the spectral flux density, $\sim 0.30 \times 10^{-17}\ \text{erg s}^{-1}\ \text{cm}^{-2}\ \AA^{-1}$ above zero.}
\label{combSpec-noEm}
\end{figure}

%\begin{figure*}[!h]
%\centering
%\includegraphics[width=18cm]{absFitVel-55320-3974-0812-rest}
%\caption{\textbf{Add Ly$\alpha$, \ion{Si}{II} - 1260, and \ion{C}{II}. Potentially find a better example.}}
%\label{metFits}
%\end{figure*}

\begin{figure}
\centering
%\resizebox{\hsize}{!}{\includegraphics[]{fitDLA-55320-3974-0812}}
\includegraphics[]{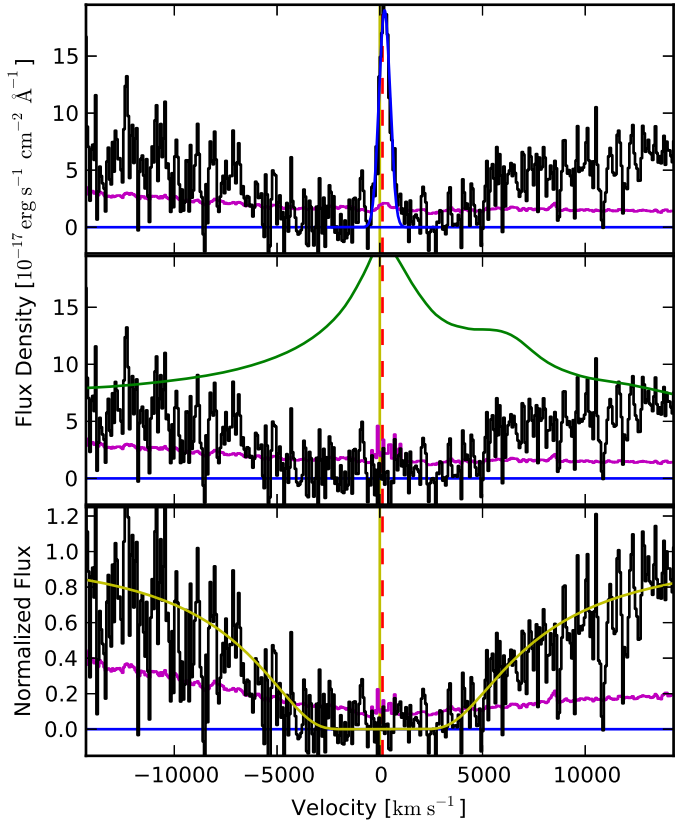}
\caption{Top: Gaussian fit (blue) to the narrow Ly$\alpha$ emission revealed by the associated DLA toward SDSS J1256+3506. Center: The narrow Ly$\alpha$ emission is subtracted, and a quasar continuum template (green) is fitted to the spectrum. Bottom: The spectrum is normalized and the \ion{H}{I} column density is fitted (yellow). SDSS J1256+3506 has log $N(\ion{H}{I}) = 22.1$. The dashed red line marks the DLA redshift, and the solid vertical yellow line indicates the velocity offset for the quasar.}
\label{emProcess}
\end{figure}

\begin{figure}
\centering
\includegraphics{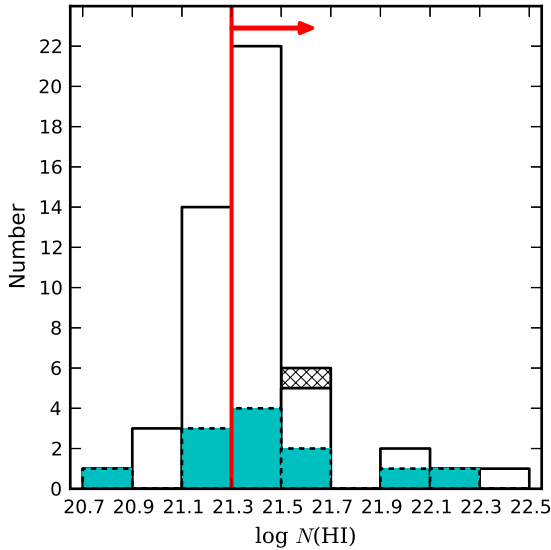}
\caption{\ion{H}{I} column density distribution for re-measured DR9 DLAs initially identified with log~$N(\ion{H}{I}) \geq 21.2$ and $\Delta v \leq 3\,000\ \rm km\ s^{-1}$. Cyan shading indicates DLAs that reveal narrow Ly$\alpha$ emission. The sample is complete above log $N(\ion{H}{i}) = 21.3$ (red arrow), and systems with lower column densities are not included in the statistical sample. One log~$N(\ion{H}{I}) \geq 21.3$ DLA (marked with cross-hatching) is also not included in the statistical sample because of the velocity difference between the DLA and the quasar.}
\label{NHIdist}
\end{figure}

The zero-level in BOSS spectra tends to be slightly higher than the true zero \citep{2012A&A...548A..66P}; this can impact the column density measurement for DLAs (Figure \ref{combSpec-noEm}). Using the Voigt profile fit to the DLA, we identify the zero-flux density region in the DLA trough and calculate the average spectral flux density across these pixels to determine the zero-level offset for each spectrum. We correct the observed spectral flux density with this offset, ensuring that the spectral flux density is truly at zero in the DLA trough.

%(Figure \ref{metFits})
Column densities were re-measured for all DLAs according to the following process. Associated low-ionization transitions redwards of Ly$\alpha$ are identified and fitted as an absorption system to obtain the DLA redshift. For all absorption line fits we use the VPFIT package\footnote{Carswell \url{http://www.ast.cam.ac.uk/~rfc/vpfit.html}}. If narrow Ly$\alpha$ emission is present in the DLA, the peak is fitted with a Gaussian profile and removed (Figure \ref{emProcess}, top). A template spectrum from the mean principal component analysis (PCA) quasar continuum \citep{2011A&A...530A..50P} is adjusted with a power-law fit to each spectrum and scaled to provide the quasar continuum near the  Ly$\alpha$ emission line (Figure \ref{emProcess}, center). The column density is measured from the normalized spectrum by fitting the DLA with the redshift fixed to the value measured from the metals (Figure \ref{emProcess}, bottom). Column density errors are on the order of 0.1 dex. Figure \ref{NHIdist} shows the resultant column density distribution in our sample. The absorption system redshift and the \ion{H}{I} column density are listed in Table \ref{DLAtable}.

\begin{figure}
%\resizebox{\hsize}{!}{\includegraphics{noiseDist}}
%\resizebox{\hsize}{!}{\includegraphics{noiseDist-zeroCorr}}
\centering
\includegraphics{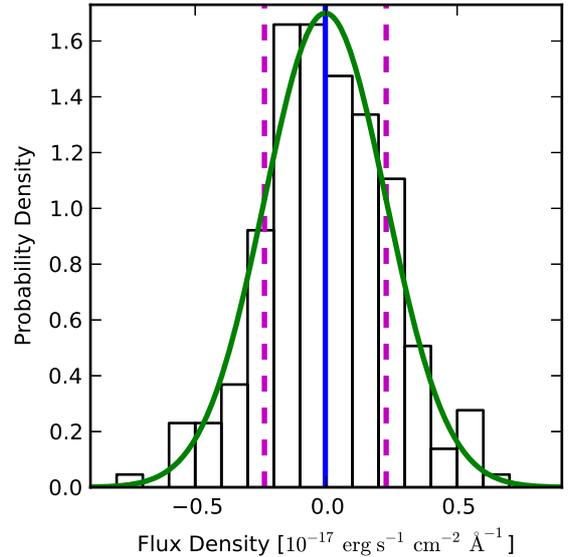}
\caption{Spectral flux density distribution for pixels in the trough of five DLAs with $\langle \text{Noise} \rangle = 0.233  \times  \rm 10^{-17}\ erg\ s^{-1}\ cm^{-2}\ \AA^{-1}$. A Gaussian fit to the spectral flux density distribution (solid green line) gives $\sigma = 0.232 \times \rm 10^{-17}\ erg\ s^{-1}\ cm^{-2}\ \AA^{-1}$ (dashed magenta lines), which is in excellent agreement with the measured average noise. The zero-level spectral flux density correction has been applied to these DLAs. }
%The center of the Gaussian (blue solid line) is at $0.065 \times  \rm 10^{-17}\ erg\ s^{-1}\ cm^{-2}\ \AA^{-1}$.
\label{charNoise}
\end{figure}

The noise level in the spectrum is the 1-$\sigma$ error on the spectral flux density. We have checked that the noise level is consistent with the scatter in the bottom of the DLA troughs by examining five DLAs with noise levels $\sigma < 0.3 \times 10^{-17}\ \rm erg\ s^{-1}\ cm^{-2}\ \AA^{-1}$ where narrow Ly$\alpha$ emission is not detected (Figure \ref{charNoise}). The average noise level for 217 pixels from the five DLAs is $\sigma = 0.233 \times 10^{-17}\ \rm erg\ s^{-1}\ cm^{-2}\ \AA^{-1}$ with $0.030 \times 10^{-17}\ \rm erg\ s^{-1}\ cm^{-2}\ \AA^{-1}$ standard deviation. The BOSS pixel size increases logarithmically from 0.82 to 2.39~\AA\ over the wavelength range 3610 -- 10\,140~\AA\ \citep{2012A&A...548A..66P}, such that the pixel size is $\sim$1~\AA\ at the typical observed wavelength for our DLAs, $\sim$4000~\AA . The $\sigma$ value from the Gaussian fit to the spectral flux density distribution, $0.232 \times 10^{-17}\ \text{erg s}^{-1}\ \text{cm}^{-2}\ \AA^{-1}$, is in excellent agreement with the average observed noise level.

We define the n-$\sigma$ detection level for the narrow Ly$\alpha$ emission (prior to its removal) as:
\begin{equation}
	\text{n} = \frac{\langle{f}_{\rm\lambda}\rangle \sqrt{\text{N}_{\text{pix}}} }{\langle\text{Err}\rangle}
\end{equation}
where $\text{N}_{\text{pix}}$ is the number of pixels within 3-$\sigma$ (1.27 FWHM) of the Gaussian fit. The average spectral flux density, $\langle{f}_{\rm\lambda}\rangle$, and average error,  $\langle\text{Err}\rangle$, are calculated for the pixels within this range. While we can determine Gaussian fits for narrow Ly$\alpha$ emission down to the $\sim$2-$\sigma$ level, we are confident about emission features detected at and above 4-$\sigma$. The DLAs that have $\geq4$-$\sigma$ narrow Ly$\alpha$ emission detections are shaded in cyan in Figure \ref{NHIdist}.

\subsection{The Statistical Sample}
The typical 1-$\sigma$ column density error in the N12 catalog is 0.20~dex. Since our more precise column density estimates do not differ from the automatic detections by more than this error, we consider that our associated DLA sample is complete for systems with log$\ N(\ion{H}{i}) \geq 21.3$. The shape of the column density distribution (Figure \ref{NHIdist}) further justifies this cut-off. %Although we searched the N12 catalog for DLAs with $\Delta v = 3\,000 \text{ km s}^{-1}$, 
We impose a velocity difference limit of $\Delta v = 1\,500 \text{ km s}^{-1}$ (stricter than the initial search criteria), which excludes one DLA with log$\ N(\ion{H}{i}) \geq 21.3$ that is 2\,530~$\text{km s}^{-1}$ below the PCA quasar redshift. (See section 4.2 for a discussion of kinematics.) The statistical sample includes 31 DLAs that are within 1\,500 $\text{km s}^{-1}$ of the quasar and have log$\ N(\ion{H}{i}) \geq 21.3$. 

The properties of these DLAs can be used to statistically look for differences between the systems that reveal narrow Ly$\alpha$ emission (detected at the 4-$\sigma$ level, 8 DLAs) and those that do not.

\subsection{The Emission Properties Sample}
\label{Sect:EmPropSamp}
To characterize the properties of the narrow Ly$\alpha$ emission features, we examine all lines-of-sight for which the emission is detected at 4-$\sigma$, regardless of the DLA column density. We supplement the DR9 sample with additional cases of DLAs revealing narrow Ly$\alpha$ emission identified during the visual inspection of quasars for DR10 \citep{2013arXiv1307.7735A}. 
%These additional spectra should become public in July 2013. 
Twenty-six DLAs are included in the emission sample. Although this dataset is not complete, it is currently the best possible sample for studying the narrow Ly$\alpha$ emission properties.

\subsection{The Redshift Distribution}
\begin{figure}
%\resizebox{\hsize}{!}{\includegraphics{zPCAdist-dr9}}
%\resizebox{\hsize}{!}{\includegraphics{zPCAcomp}}
\centering
\includegraphics{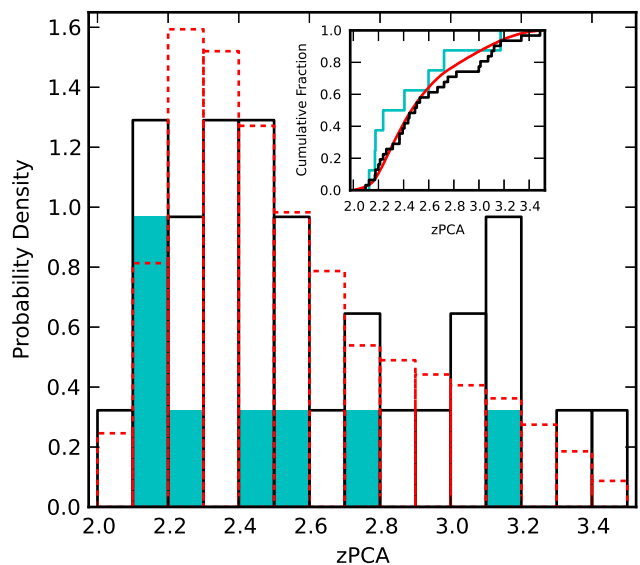}
\caption{Distribution of quasar redshifts estimated from a principal component analysis for the statistical sample (solid black line) and the DR9 quasars (dashed red line). The relative portion of DLAs in the statistical sample that reveal narrow Ly$\alpha$ emission are shaded in cyan. The inset gives the cumulative fraction function using the same color scheme.}
%\label{zPCAdist-dr9}
\label{zPCAcomp}
\end{figure}

Figure \ref{zPCAcomp} compares the redshift distributions for quasars in the statistical sample and DR9. %across the range of redshifts observed in the complete sample.
The most noticeable difference is a possible excess of quasars at $z = 3.0 - 3.2$ in the statistical sample. The peak in probability density is also flatter for the statistical sample than for the DR9 quasars. Otherwise, the redshift distribution for quasars in the statistical sample generally follows that of the DR9 quasars. Based on a Kolmogorov-Smirnov test, it is 68\% likely that the two distributions are the same. A K-S test comparing the eight quasars in the statistical sample that have detected narrow Ly$\alpha$ emission with the DR9 quasars indicates that it is 26\% likely the two populations come from the same distribution. The quasar redshift distribution also shows that there is no preferred redshift for DLAs that reveal narrow Ly$\alpha$ emission. 

% \ion{O}{I} - 1302\AA, \ion{Si}{II} - 1260, 1304, 1526\AA, \ion{C}{II} - 1334\AA, \ion{Al}{II} - 1670\AA, \ion{Fe}{II} - 2344, 2374, 2382, 2586, 2600\AA, \ion{Mg}{II} - 2796, 2803\AA, and \ion{Mg}{I} - 2852\AA.

%
%______________________________________________________________
\section{Anticipated Number of Intervening DLA Systems within $1\,500\ \rm km\ s^{-1}$ of zQSO}
We explore the possibility of an overdensity of DLAs at the quasar redshift by calculating the anticipated number of intervening DLAs with $\Delta v \leq 1\,500\ \rm km\ s^{-1}$ in DR9 and comparing this result to the number observed in our statistical sample of associated DLAs.

\subsection{Anticipated Number in DR9}
The absorption path is given by:
\begin{equation}
	\frac{d\chi}{dz} = \frac{(1+z)^{2}}{\Omega_{\text{m}}(1+z)^{3}+\Omega_{\text{k}}(1+z)^{2}+\Omega_{\Lambda}}
\end{equation}
%  to get $d\chi \approx \frac{1}{12}$
\citep{1969ApJ...156L...7B} where $dz = (1+z)\frac{\Delta v}{c}$. We evaluate this expression with $\Delta v = 1\,500 \text{ km s}^{-1}$ and the PCA redshift for each non-BAL quasar in DR9 with $2.0 \leq z_{\rm PCA} \leq 3.5$, the redshift range covered in the statistical sample. Summing the absorption path for each of the 55\,679 quasar lines-of-sight gives the total absorption path, $\Delta\chi$.

We use the $N(\ion{H}{I})$ distribution function, $f(N_{\ion{H}{I}}, \chi)$, given in N12 to integrate
\begin{equation}
	\frac{N_{\text{DLA}}}{d\chi} = \int f(N_{\ion{H}{I}}, \chi)\, d N_{\text{QSO}}
\end{equation}
for $N_{\ion{H}{I}} \geq 10^{21.3} \text{cm}^{-2}$, since our statistical sample is complete above this column density. The expected number of DLAs is then $N_{\text{DLA}} = \frac{N_{\text{DLA}}}{d\chi} \Delta \chi$, which predicts 12.93 associated DLAs in DR9.

\subsection{The Effect of Clustering near Quasars}
If the number density of associated DLAs is the same as that of intervening DLAs, then we anticipate finding $\sim$13 associated DLAs with log $N(\ion{H}{I}) \geq 21.3$ in DR9. Our statistical sample contains 31 such DLAs, more than twice as many as expected. This observed overdensity can be attributed to the clustering properties of DLAs around quasars. 
It is well known that quasars occur in overdense regions. Measurements of quasar clustering, which is more pronounced at higher redshifts \citep{2005MNRAS.356..415C}, indicate that a quasar is likely to be part of a group of galaxies. It has also been shown that an overdensity of neutral gas is present within $\sim$10~Mpc of the quasar \citep{2007MNRAS.377..657G, 2013arXiv1303.1937F}.
% where the probability of finding high density gas either around the host galaxy itself or in nearby galaxies is higher. 
%From clustering arguments, (Following from these arguments) 
Not surprisingly, the incidence of DLAs within 3\,000~$\text{km s}^{-1}$ of the quasar is $\sim$2--4 times higher than that of intervening DLAs \citep{2010MNRAS.406.1435E}. 
The number of associated DLAs that we observe is therefore consistent with what we expect due to clustering around the quasar host galaxy.

%
%______________________________________________________________
\section{Characterizing the Statistical Sample}

\subsection{Scenario}

The statistical sample breaks into two populations: associated DLAs that reveal narrow Ly$\alpha$ emission at more than the 4-$\sigma$ level (8) and those with no emission detected at this level (23). The first category is such an unusual occurrence that we investigate whether these DLAs have specific properties.

We would like to test the hypothesis that the absorbers giving rise to associated DLAs with narrow Ly$\alpha$ emission superimposed on their troughs are intrinsically different than the absorbers responsible for the associated DLAs where no emission is detected. Narrow Ly$\alpha$ emission appears in the DLA trough only if the absorber blocks the $\sim$1~pc BLR without also covering more extended sources of Ly$\alpha$ emission, such as the NLR or star-forming regions. When no Ly$\alpha$ emission is observed, a galaxy near to the quasar host is likely acting as a screen, blocking the BLR, NLR, and any extended emission regions. 

The impact parameters for ten confirmed $z \gtrsim 2$ galaxy counterparts to intervening DLAs are all on the order of ten kiloparsecs \citep{2012MNRAS.424L...1K}. For very high column densities, such as those observed in our sample, impact parameters could be even smaller \citep[e.g.][]{2012A&A...540A..63N}. A neighboring galaxy would be sufficiently extended to obscure Ly$\alpha$ emission from the quasar host, but the absorber that allows narrow Ly$\alpha$ emission to pass unobscured must have a more compact size.
\citet{2009ApJ...693L..49H} suggest that the associated DLA towards SDSS J1240+1455 with a narrow Ly$\alpha$ emission peak could arise from small ($\sim$10~pc), dense ($\rm \sc{n}_{H}$~$\sim$~100~$\rm cm^{-3}$) clouds of neutral hydrogen in the quasar host galaxy.
Further supporting the idea that not all associated absorptions are due to galaxies clustered near the quasar host, \citet{2008MNRAS.388..227W} found that upwards of 40\% of 1.6 < z < 4 $\Delta v \leq$ 3\,000 km $\rm s^{-1}$ \ion{C}{IV} absorbers are directly related to the quasar. We likewise propose that the DLAs revealing narrow Ly$\alpha$ emission are due to dense clouds of neutral hydrogen in the quasar host galaxy.

\subsection{Kinematics}

\begin{figure}[!ht]
\centering
\includegraphics{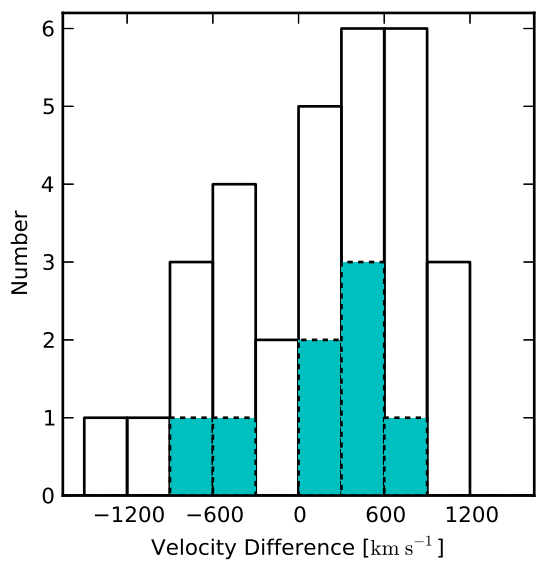}
\caption{Velocity difference distribution for $z_{\rm \, DLA}$ with respect to $z_{\rm \, QSO}$. DLAs that reveal narrow Ly$\alpha$ emission are shaded in cyan. The preference for positive velocity differences may arise from uncertain quasar redshift estimates, rather than an abundance of DLAs falling into the quasar.} %A dashed line indicates the average velocity difference.
\label{velDist-dr9}
\end{figure}

The velocity difference of the DLA redshift, measured from the metals, with respect to the quasar redshift, determined from a PCA, is not distinct for systems with and without superimposed narrow Ly$\alpha$ emission (Figure \ref{velDist-dr9}). For both populations, nearly all DLAs fall within 1\,200 $\text{km s}^{-1}$ of the quasar. 

The main uncertainty when studying the kinematics is that the quasar redshift is not well known. More than half of the DLAs appear to be at higher redshifts than the quasars (systems with positive velocity difference in Figure \ref{velDist-dr9}), which could indicate infalling absorbers. However, positive velocity differences are consistent with a known tendency for the PCA redshifts to be slightly bluer than the true quasar redshift \citep{2012A&A...548A..66P}.

\subsection{Metals}

\begin{figure}
\centering
%\resizebox{\hsize}{!}{\includegraphics{ks-W0}}
\includegraphics{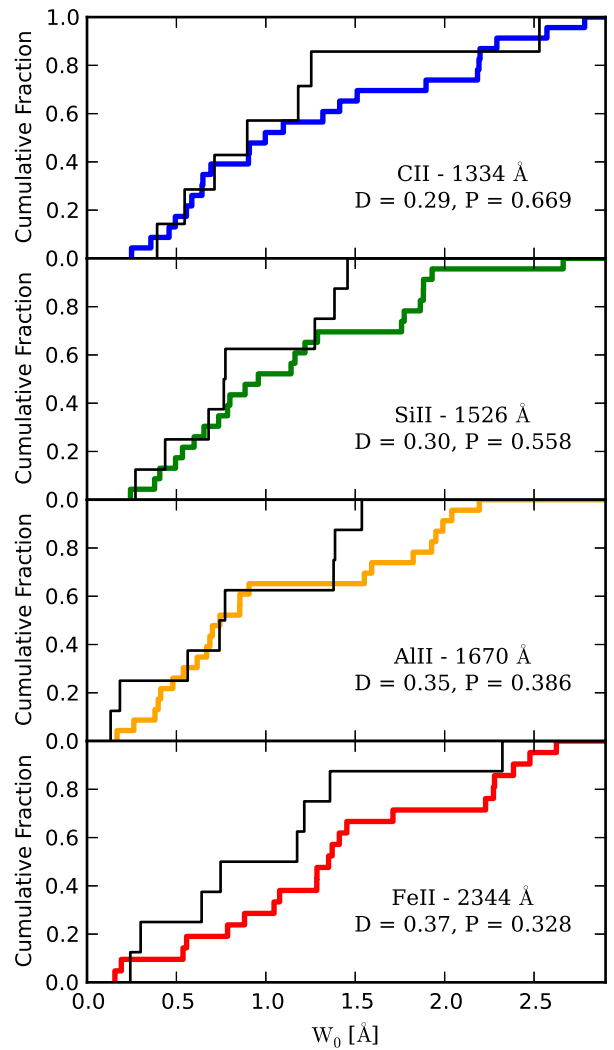}
\caption{Kolmogorov-Smirnov tests comparing the DLAs with and without detected narrow Ly$\alpha$ emission using the rest equivalent widths for \ion{C}{II}~-~1334~\AA, \ion{Si}{II}~-~1526~\AA, \ion{Al}{II}~-~1670~\AA, and \ion{Fe}{II}~-~2344~\AA\ absorption features. A black line traces the $\rm W_{0}$ distribution for DLAs with narrow Ly$\alpha$ emission detected, and a thick colored line indicates the $\rm W_{0}$ distribution for those without an emission detection. The K-S test statistic, D, gives the maximum vertical distance between the two distributions, and the P-value is the probability that the two distributions are drawn from the same population.}
\label{ksTests}
\end{figure}

\begin{figure}[!h]
\centering
%\resizebox{\hsize}{!}{\includegraphics{FeII-CII-depl}}
\includegraphics{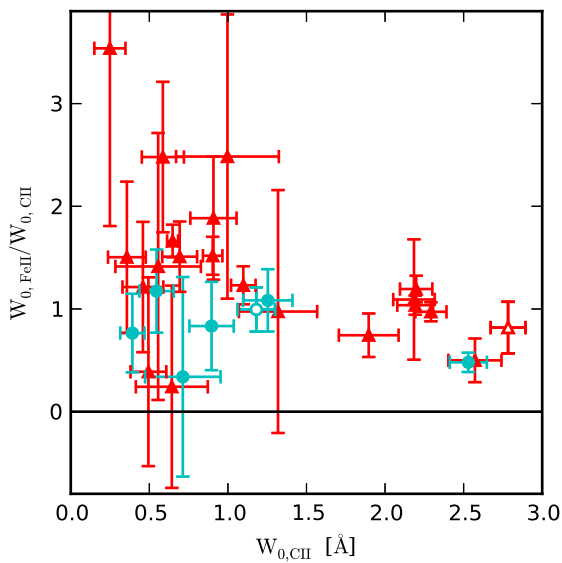}
\caption{ The equivalent width ratio $\left. {W_{0,\ \ion{Fe}{II}}} \middle/ {W_{0,\ \ion{C}{II}}} \right.$, is plotted as a function of $W_{0,\ \ion{C}{II}}$. DLAs where narrow Ly$\alpha$ is detected are marked by blue circles, and DLAs without a 4-$\sigma$ detection are marked by red triangles. Filled symbols indicate that $W_{0}$ is measured from the fit to the absorption line, and empty symbols indicate that $W_{0}$ is estimated from the flux. One spectrum has flux problems at the location of the \ion{C}{II}~-~1334~\AA\ absorption, and \ion{Fe}{II}~-~2344~\AA\ is redshifted out of the spectrum in two cases. Twenty-nine of the 31 DLAs in the statistical sample are plotted here, 7 with an emission detection and 22 without.}
\label{csDepletion}
\end{figure}

\begin{figure}
\centering
\resizebox{0.9\hsize}{!}{\includegraphics[bb=100 182 315 570,width=0.5\hsize]{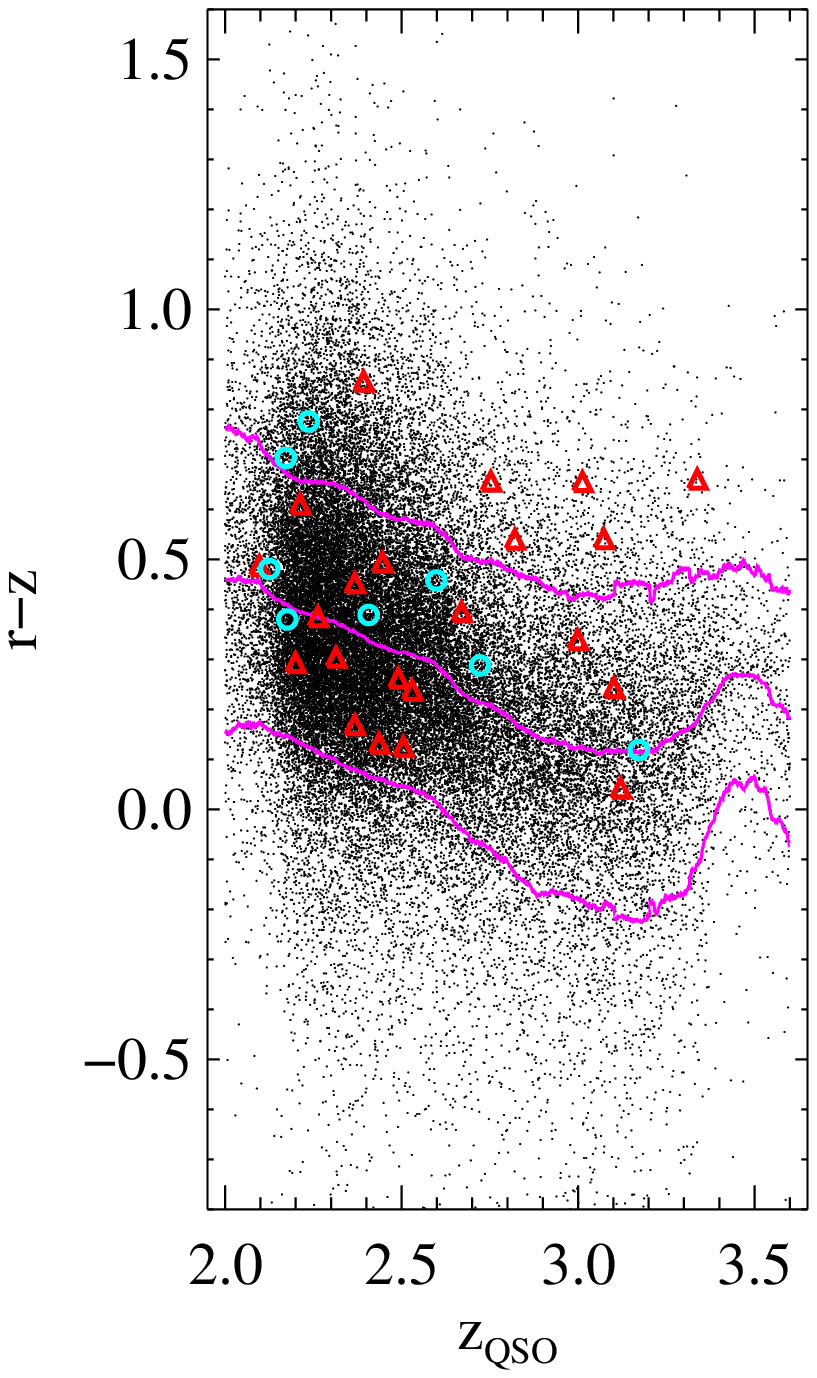} \includegraphics[bb=100 182 315 570,width=0.5\hsize]{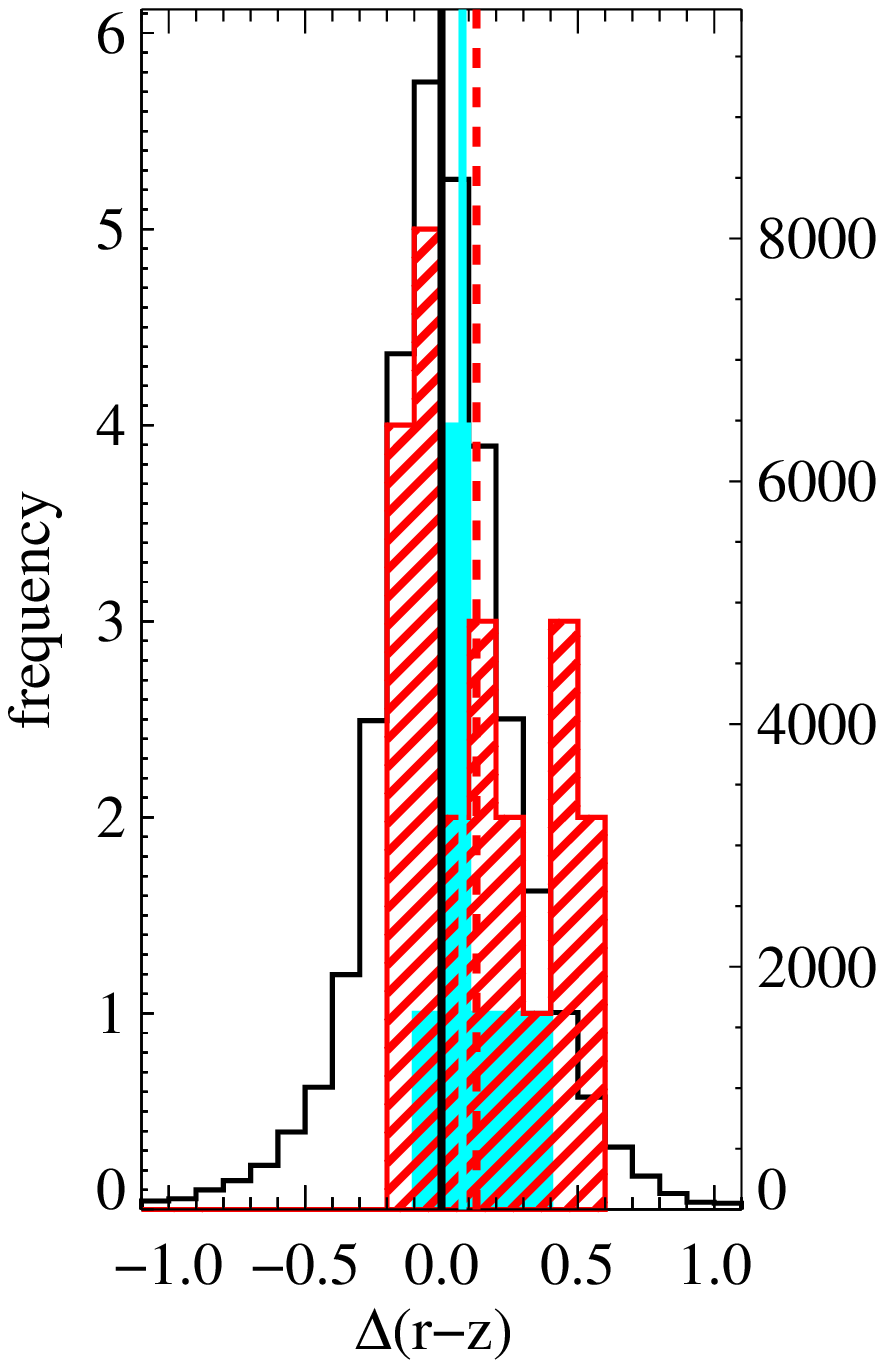}}
%\resizebox{0.5\hsize}{!}{\includegraphics[bb=75 182 315 570,width=0.5\hsize]{hist_rz}}
\caption{The $r$-$z$ color as a function of the quasar redshift (left) and the $r$-$z$ color distribution (right) for DR9 quasars without BALs or a DLA along the line-of-sight (black), associated DLAs without an emission detection (red), and associated DLAs with narrow Ly$\alpha$ emission detected (cyan). In the color plot, the median and 1-$\sigma$ colors for the DR9 quasar sample are shown as magenta curves.}
\label{colors}
\end{figure}

\begin{figure}
\centering
%\resizebox{\hsize}{!}{\includegraphics{Eb-v_dist}}
\includegraphics{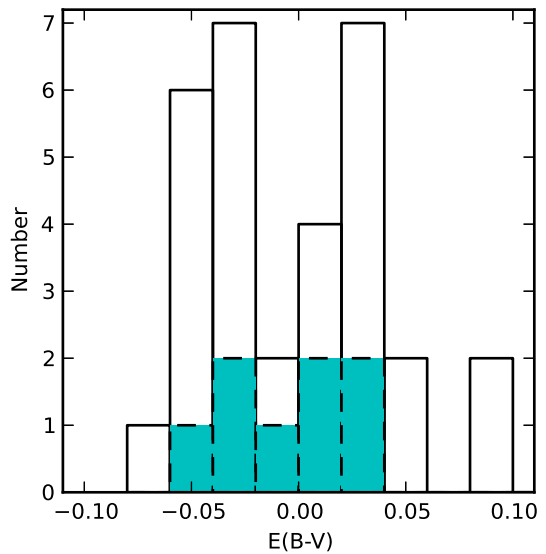}
\caption{E(B-V) reddening distribution estimated from a Small Magellanic Cloud reddening law template. DLAs with narrow Ly$\alpha$ emission detected are shaded in cyan.}
\label{csReddening}
\end{figure}

The rest equivalent width, $\text{W}_{0}$, of various metal transitions, \ion{C}{II}~-~1334~\AA, \ion{Si}{II}~-~1526~\AA, \ion{Al}{II}~-~1670~\AA, and \ion{Fe}{II}~-~2344~\AA, are measured for each DLA in the sample. These absorption features were chosen because they are typically not blended or redshifted out of the spectrum. The signal-to-noise ratio (S/N) is also best at wavelengths redward of Ly$\alpha$. However, sky subtraction problems start to become noticeable after 7500 \AA, and in some spectra the \ion{Fe}{II} lines are subject to additional noise.

Table \ref{W0table} gives $\text{W}_{0}$ values measured from a fixed-redshift Voigt profile fit to the absorption system (referred to as "Fit" in the Table). A second estimate of $\text{W}_{0}$ is directly measured from the normalized flux by integrating over the line profile (referred to as "\spef" in the Table). We include both estimates of $\text{W}_{0}$ because of local problems in the spectra and/or complex blends that are difficult to disentangle at the spectral resolution and S/N typical of BOSS data. The error on $\text{W}_{0}$ is calculated using the error on the normalized flux across the six innermost pixels.

We performed a two-distribution K-S test with $\text{W}_{0}$ for the populations where emission is and is not detected (Figure \ref{ksTests}). In general, the largest $\text{W}_{0}$ values are observed in systems where no emission is detected. However, the \ion{C}{II}~-~1334~\AA\ and \ion{Fe}{II}~-~2344~\AA\ distributions each have a strong absorption in a DLA system with emission detected. Overall, the $\text{W}_{0}$ values are less than $\sim$67\% likely to come from the same population. The maximum distance parameter is largest for \ion{Fe}{II}~-~2344~\AA , and for this absorption feature the probability that the populations are the same is only 33\%. 

Since most of these absorption features are saturated, it is reasonable to assume that $W_{0}$ is related to the width of the lines and not to column densities. Keeping this in mind, we nonetheless plot $\left. {W_{0,\ \ion{Fe}{II}}} \middle/ {W_{0,\ \ion{C}{II}}} \right.$ versus $W_{0,\ \ion{C}{II}}$ (Figure \ref{csDepletion}) to investigate any indication of species depleting onto dust grains. While some depletion is expected for \ion{C}{II}, \ion{Fe}{II} is expected to be $\sim$50 times more depleted \citep{1999ApJS..124..465W}. The equivalent width ratios appear to decrease with increasing $W_{0,\ \ion{C}{II}}$. A linear fit indicates a slope of $-0.39 \pm 0.07\ \AA^{-1}$ and a y-intercept of $1.74 \pm 0.15$. The systems where narrow Ly$\alpha$ emission is detected consistently have lower $\left. {W_{0,\ \ion{Fe}{II}}} \middle/ {W_{0,\ \ion{C}{II}}} \right.$ than the systems without a detection at the same $W_{0,\ \ion{C}{II}}$. 
We should be careful when interpreting this, but if the $\left. {W_{0,\ \ion{Fe}{II}}} \middle/ {W_{0,\ \ion{C}{II}}} \right.$ is related to depletion then this would indicate that dust content is
stronger in DLAs showing emission. Although somewhat surprising, this would at least support the idea that the gas in the DLA is not directly related to the Ly$\alpha$ emitting gas. 
Any correlations may become clearer with column densities, rather than $W_{0}$, but the BOSS spectral resolution and S/N are too low to permit this measurement.

\subsection{Reddening}

We explore reddening of the quasar spectra due to dust absorption from the associated DLAs by comparing their colors to the general population of quasars. All but one of the associated DLAs fall in the $u$ or $g$ band, so we are limited to the $r$-$i$ and $r$-$z$ colors. The left panel of Figure \ref{colors} shows the $r$-$z$ color as a function of $z_{\rm \, QSO}$ and the right panel gives the distribution for color bins. Quasar spectra with an associated DLA and narrow Ly$\alpha$ emission are nearly all redder than the average $r$-$z$ color for DR9 quasars with no DLA along their line-of-sight, while the $r$-$z$ colors for associated DLAs without a narrow Ly$\alpha$ detection are more dispersed. The distribution reveals that the associated DLAs, both with and without an emission detection, have redder average $r$-$z$ colors than the DR9 comparison quasars. A K-S test indicates that the associated DLAs with (without) narrow Ly$\alpha$ emission detected have a  3\% (36\%) probability of arising from the same $r$-$z$ color distribution as the DR9 comparison quasars, and the associated DLAs with an emission detection have a 21\% probability of arising from the same distribution as the associated DLAs without a detection. A larger sample size would be needed to confirm that these are truly distinct populations.

We fit the quasar spectra with a Small Magellanic Cloud (SMC) reddening law template \citep{2003ApJ...594..279G} to further investigate potential differences between the associated DLAs that reveal narrow Ly$\alpha$ emission and those that do not.
%\textbf{The SMC , rather than a Large Magellanic Cloud or Milky Way extinction curve, is the best fit to the quasar spectra in the majority of cases, because of the lack of 2175 \AA\ bump in the spectra.}
{The SMC extinction curve is the best fit for the majority of the spectra, because they lack the 2175 \AA\ bump characteristic of Large Magellanic Cloud or Milky Way extinction curves.}
 In the E(B-V) reddening distribution (Figure \ref{csReddening}), the most reddened systems (E(B-V) $> 0.05$) do not have a narrow Ly$\alpha$ emission detection. Otherwise, the distributions are similar for the two populations, and they are both consistent with no reddening on average.

%
%______________________________________________________________

\section{Characterizing the Narrow Ly$\alpha$ Emission}

In order to explore the properties of the narrow Ly$\alpha$ emission detected in the troughs of associated DLAs, we analyze the sample described in Section \ref{Sect:EmPropSamp}.
%The eight DLAs in the statistical sample with detected Ly$\alpha$ emission are included in the emission properties sample (26 DLAs total).  

%We derive $z_{\rm\, Ly\alpha}$, the integrated flux (IF), and the luminosity from a Gaussian fit to the narrow Ly$\alpha$ emission.

%in the narrow Ly$\alpha$ emission peak 
We fit a Gaussian profile to each narrow Ly$\alpha$ emission peak and derive $z_{\rm\, Ly\alpha}$, the integrated flux (IF), and the luminosity from the fit parameters. The integrated flux is calculated by summing the spectral flux density in the region within 3-$\sigma$ of the Gaussian fit center (1.27 FWHM). The emission peaks all have IF values less than $70 \times 10^{-17}\ \rm erg\ s^{-1}\ cm^{-2}$, except for two extremely strong emission peaks with IF $\sim$200~$\rm \times 10^{-17}\ erg\ s^{-1}\ cm^{-2}$. Since no IF values are in the range $\rm 26 < IF < 40 \times 10^{-17}\ erg\ s^{-1}\ cm^{-2}$, we use these limits to distinguish strong and moderate narrow Ly$\alpha$ emission peaks.

%We obtain a luminosity distance from $z_{\rm\, Ly\alpha}$ and combine this with the integrated flux to get the luminosity. 

We estimate the luminosity using the integrated flux and the luminosity distance calculated for $z_{\rm\, Ly\alpha}$. In Figure \ref{setOf3}, we examine properties of the DLA, quasar, and emission peak as a function of the integrated flux and the luminosity.

\subsection{Correlation with other properties}
 
\begin{figure*}[!phtb]
\centering
\includegraphics[scale=0.97]{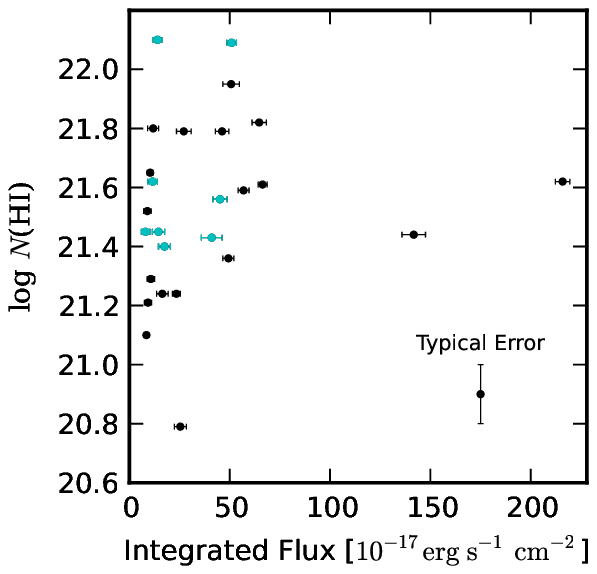}
\includegraphics[scale=0.97]{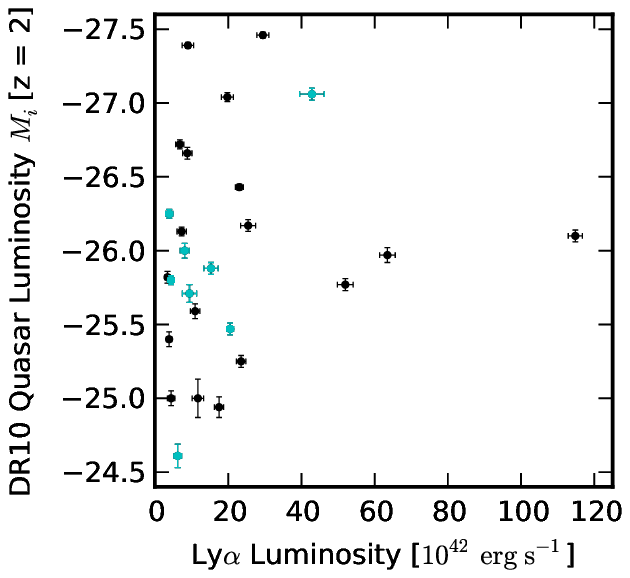}
\includegraphics[scale=0.97]{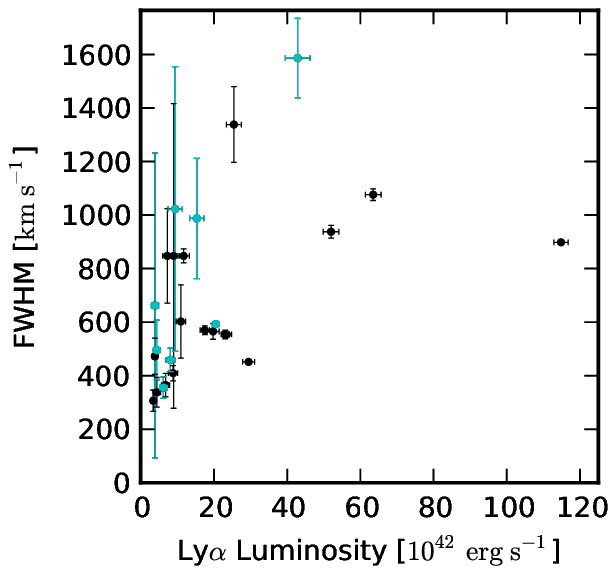}
\caption{ \ion{H}{I} column density as a function of the integrated flux in the narrow Ly$\alpha$ emission peak (left), the quasar absolute luminosity as a function of the Ly$\alpha$ emission luminosity (center), and the deconvolved FHWM from a Gaussian fit to the narrow Ly$\alpha$ emission peak (right) as a function of the Ly$\alpha$ emission luminosity. DLAs that are part of the statistical sample are shaded in cyan.}
\label{setOf3}

\end{figure*}

The \ion{H}{I} column density increases mildly with integrated flux (Figure \ref{setOf3}, left). 
An outlier with log $N(\ion{H}{I}) \approx 20.8 $ has moderate Ly$\alpha$ emission, and none of the log $N(\ion{H}{I}) < 21.3 $ DLAs show strong Ly$\alpha$ emission. 
%\textit{What to say about $W_{0,\ \ion{C}{II}}$? (Figure \ref{4vsIF}, upper right)}
\newline 
\noindent
The quasar absolute luminosity, indicated by $M_{i}\ [z = 2]$, is calculated using the $K$-correction process outlined in \citet{2006AJ....131.2766R}, with $\alpha_{\nu} = -0.5$ and H0 $= 70\rm\ km s^{-1} Mpc^{-1}$. No trend is apparent for the quasar absolute luminosity versus Ly$\alpha$ emission luminosity (Figure \ref{setOf3}, center).
\newline 
\noindent
The FWHM from the Gaussian fit to the narrow Ly$\alpha$ emission peak is deconvolved from the BOSS spectrograph velocity resolution, 150~km~$\rm s^{-1}$, and the error depends on the error for the fit to the Gaussian $\sigma$ parameter. Two narrow Ly$\alpha$ emission peaks have FWHM $> 1\,100\ \rm km\ s^{-1}$. Only one point at FWHM~$= 662\ \rm km\ s^{-1}$ with a large uncertainty falls between $\sim 600 \ \rm km\ s^{-1}$ and $\sim 850 \ \rm km\ s^{-1}$.  

\begin{figure}[!phtb]
\centering
\includegraphics{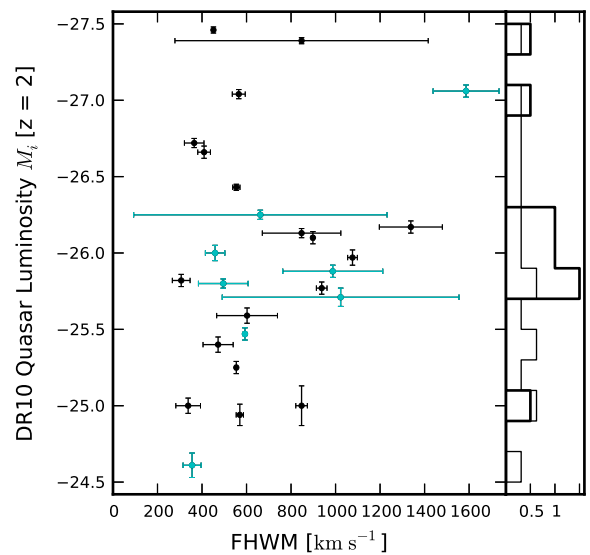}
\caption{Quasar absolute luminosity as a function of the deconvolved FWHM from the Gaussian fit to the narrow Ly$\alpha$ emission. DLAs that are part of the statistical sample are shaded in cyan. The probability density distribution for the luminosity of quasars that have narrow Ly$\alpha$ emission peaks with FWHM $\lesssim 450\ \text{km s}^{-1}$ (thin line) and FWHM $\gtrsim 600\ \text{km s}^{-1}$ (thick line) is given to the right. }
\label{LvsFWHM}
\end{figure}

Figure \ref{LvsFWHM} shows the quasar absolute luminosity as a function of the FWHM. Narrow Ly$\alpha$ emission peaks with FWHM~$\lesssim 660\ \text{km s}^{-1}$ are uniformly distributed across the luminosity range, while those with FWHM~$\gtrsim 850\ \text{km s}^{-1}$ are concentrated at $\left\langle \text{M}_{i} \right\rangle = -26.1$. The average FWHM for the narrower and wider Ly$\alpha$ emission peaks are $\sim$485~$\text{km s}^{-1}$ and $\sim$1\,000~$\text{km s}^{-1}$ respectively. For the wider Ly$\alpha$ emission peaks, luminosity increases with increasing FWHM. Ly$\alpha$ emission peaks with large FWHM values are preferentially observed in high luminosity quasars.


\begin{figure}[!htbp]
%\resizebox{\hsize}{!}{\includegraphics{velDiffEm-hist}}
\centering
\includegraphics{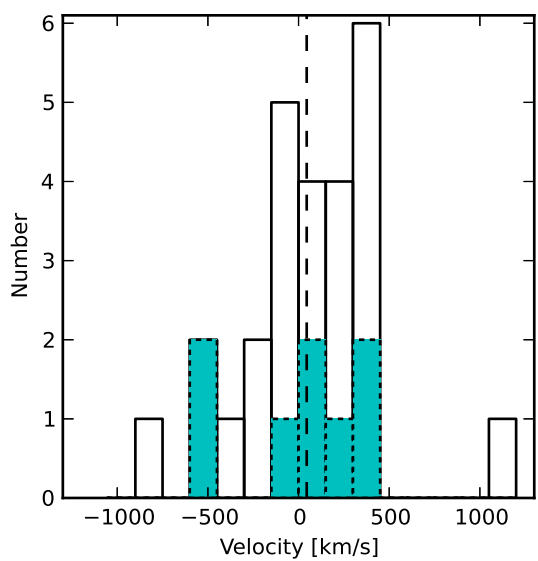}
\caption{Velocity difference distribution for $z_{\rm \, Ly\alpha}$ with respect to $z_{\rm \, DLA}$. A dashed line marks the mean velocity difference. DLAs that are part of the statistical sample are shaded in cyan.}
\label{velDiff-em}
\end{figure}

\subsection{Position and profile of the emission line}

The center of the Gaussian fit to the narrow Ly$\alpha$ emission provides the Ly$\alpha$ emission redshift, $z_{\rm\, Ly\alpha}$. The velocity difference distribution for $z_{\rm \, Ly\alpha}$ with respect to $z_{\rm \, DLA}$ reveals that the narrow Ly$\alpha$ emission is within $\pm 500\ \text{km s}^{-1}$ of the DLA for the majority of systems (Figure \ref{velDiff-em}). The narrow Ly$\alpha$ emission peaks are blueshifted and redshifted in nearly equal numbers, and the mean velocity difference is $45\ \rm km\ s^{-1}$.

\begin{figure}[!phtb]
\centering
%\resizebox{\hsize}{!}{\includegraphics[scale=0.5]{strongEmComb-only}}
\includegraphics[]{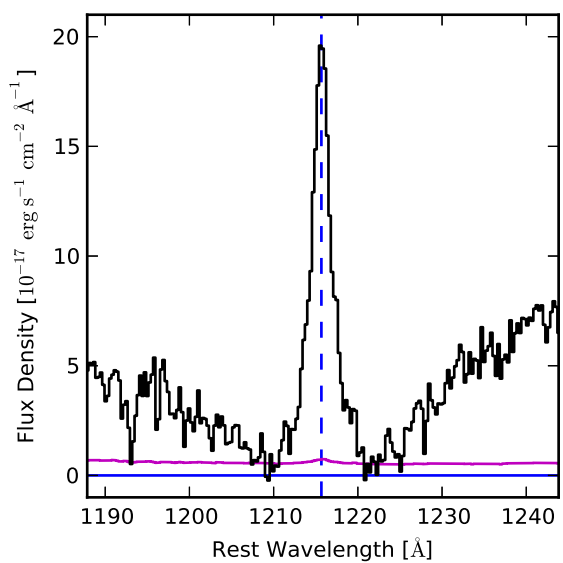}
\caption{Combined spectrum for eleven DLAs with strong Ly$\alpha$ emission (IF $> 40 \times 10^{-17}\ \rm erg\ s^{-1}\,cm^{-2}$). The spectral flux density zero-level is corrected, and the spectra are combined with an inverse variance weighted average.}
\label{combSpec-Strong}
\end{figure}

\begin{figure}[!phtb]
%\resizebox{\hsize}{!}{\includegraphics{modEmComb}}
\centering
\includegraphics[]{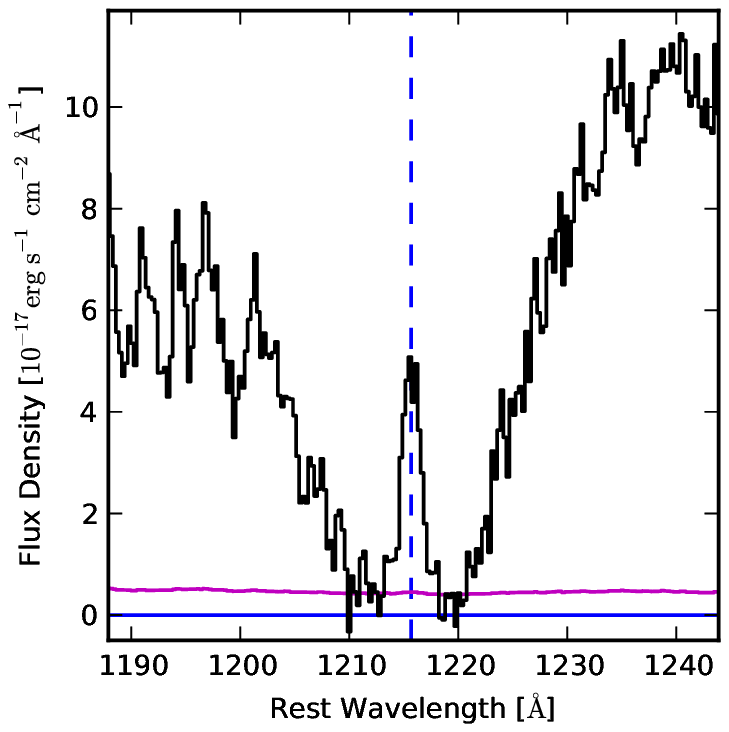}
\caption{Combined spectrum for fifteen DLAs with moderate Ly$\alpha$ emission (IF $\leq 26 \times 10^{-17} \ \rm erg\ s^{-1}\,cm^{-2}$). The spectral flux density zero-level is corrected, and the spectra are combined with an inverse variance weighted average.}
\label{combSpec-Mod}
\end{figure}

Figures \ref{combSpec-Strong} and \ref{combSpec-Mod} show the combined spectra for the associated DLAs with strong and moderate narrow Ly$\alpha$ emission, respectively. The emission is symmetric, in contrast to the asymmetric line profile characteristic of Lyman Break Galaxies \citep[LBGs;][]{2007A&A...467...63T}. This indicates that there is no \ion{H}{I} gas in front of the emitting region to absorb the emitted Ly$\alpha$ photons. Small \ion{H}{I} clouds in the host galaxy can explain the presence of the associated DLAs and also the symmetric narrow Ly$\alpha$ emission. Neutral hydrogen is located in front of the quasar line-of-sight, but does not fully cover the NLR or extended emission regions.

\begin{figure}[!phtb]
%\resizebox{\hsize}{!}{\includegraphics{combSpec_zCorr_noEm_IF6}}
\centering
\includegraphics[]{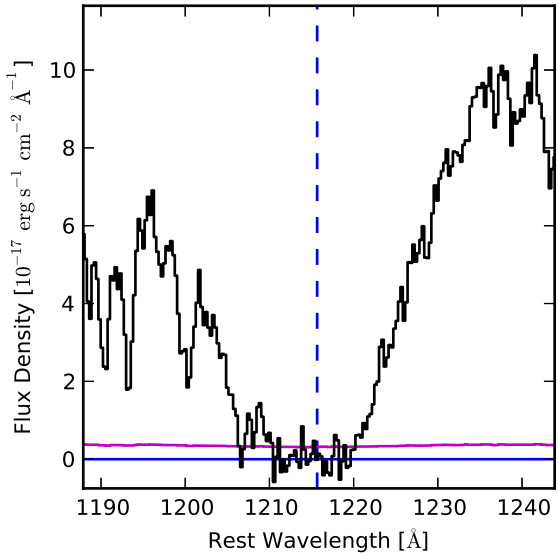}
\caption{Combined spectrum for fifteen DLAs that have an integrated Ly$\alpha$ emission line flux upper limit $\leq 5.5 \times 10^{-17}\ \rm erg\ s^{-1}\ cm^{-2}$ and $N(\ion{H}{I}) \geq 21.3$. The spectral flux density zero-level is corrected, and the spectra are combined with an inverse variance weighted average.}
\label{combSpec-noEmLim}
\end{figure}

%We have stacked all the spectra where we do not detect an individual emission.
Figure \ref{combSpec-noEmLim} presents the combined spectrum for the associated DLAs where the 4-$\sigma$ upper limit on the integrated flux is $\leq 5.5 \times 10^{-17}\ \rm erg\ s^{-1}\ cm^{-2}$. At the 3-$\sigma$ level, no Ly$\alpha$ emission is detected in the combined spectrum down to $0.79 \times 10^{-17}\ \rm erg\ cm^{-2}\ s^{-1}$, which is surprising for such bright quasars. This implies that the entire host galaxy must be covered by a screen of gas unrelated to its own composition. We speculate that an independent nearby galaxy acts as a screen and blocks the Ly$\alpha$ emission from the quasar host galaxy.

\subsection{Comparison with Lyman Break Galaxies}

The narrow Ly$\alpha$ emission peaks superimposed on the troughs of associated DLAs can be compared with LBGs, which typically have redshifted emission and blueshifted absorption features at the Ly$\alpha$ transition. In LBGs, star-forming regions that emit Ly$\alpha$ photons are surrounded by \ion{H}{I} gas. The Ly$\alpha$ photons are most likely to escape the galaxy when they scatter off of gas that is redshifted with respect to the emitting region. %have the best chance of escaping 

\citet{2003ApJ...588...65S} analyzed spectra from 811 LBGs at $z\sim3$, and found that the average velocity difference between Ly$\alpha$ emission and low ionization interstellar absorption is $650\ \text{km s}^{-1}$. The deconvolved Ly$\alpha$ emission FWHM measured from a composite spectrum of the LBG sample, $\rm 450 \pm 150\ km\ s^{-1}$, is in the range of deconvolved FWHM values measured for the narrow Ly$\alpha$ emission peaks observed in the associated DLAs. However, the FWHM values for this later category indicate two populations of Ly$\alpha$ emission peaks, one of which is comparable to and another that is more than twice as wide as the average FWHM observed in LBGs. The Ly$\alpha$ emission in LBGs also varies with UV luminosity: fainter galaxies have larger Ly$\alpha$ emission equivalent widths than brighter galaxies. No such trend is identified for the associated DLAs.

The narrow Ly$\alpha$ emission peaks revealed by associated DLAs are not systematically redshifted like those in LBGs. This, along with the symmetric emission peak profiles, is clear evidence that the emission is not significantly scattered and therefore that the emitting region is relatively void of \ion{H}{I} gas. Ly$\alpha$ emission features in LBGs and associated DLAs appear to have different properties. The absorbing gas that gives rise to an associated DLA is separate from the emission region. 

\subsection{Comparison with Lyman-Alpha Emitters}

\begin{figure}
%\resizebox{\hsize}{!}{\includegraphics{nonZeroAndInterDLA-noEm}}
\centering
\includegraphics{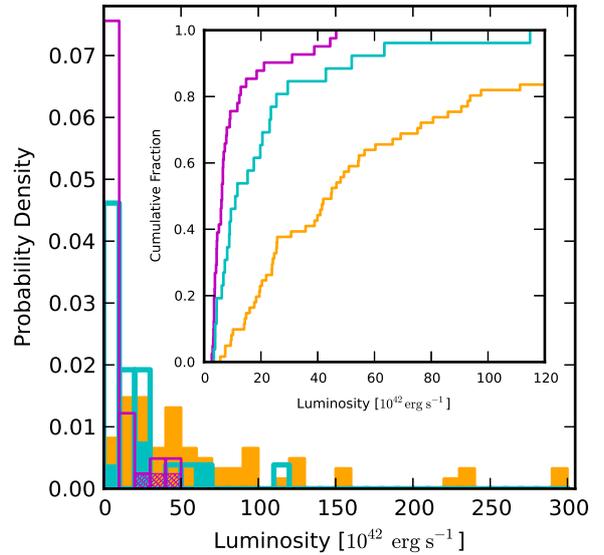}
\caption{Distribution of Ly$\alpha$ emission luminosities for our sample (thick cyan), LAEs (magenta), and radio galaxies (filled yellow). Cross-hatched magenta shading denotes LAEs that have an AGN contribution (3). Cyan shading indicates quasars in our sample that are also radio-loud (4). The inset shows the K-S test cumulative fraction out to $120 \times 10^{42}\ \rm erg\ s^{-1}$ with the same color scheme. The cumulative fraction distribution is scaled for a comparison between our sample and LAEs and is truncated for the radio galaxies.}
\label{lumHist}
\end{figure}

Lyman-alpha emitters (LAEs) are 
%another class of 
objects detected on the basis of their Ly$\alpha$ emission. Their emission is primarily due to star formation, since these galaxies typically have negligible AGN activity except at the highest luminosities \citep[log~$L>43.5$~erg~$\rm s^{-1}$, see][]{2008ApJS..176..301O, 2013ApJ...771...89O}. We therefore expect the luminosities for our sample of narrow Ly$\alpha$ emission peaks detected by using DLAs as coronagraphs to be consistently higher than what is observed in LAEs, due to the AGN ionizing gas in its host galaxy. 
Figure \ref{lumHist} compares the luminosities of our narrow Ly$\alpha$ emission peaks and a complete sample of 41 $z\sim3.1$ LAEs from \citet{2008ApJS..176..301O}, three of which have an AGN contribution. 
The majority of the LAEs have a luminosity less than $10^{43}\ \rm erg\ s^{-1}$. 
%Fewer than a quarter (10) of the LAEs have a luminosity greater than $10^{43} \rm ergs\ s^{-1}$, whereas over half (14) of the Ly$\alpha$ emission peaks in our sample have luminosities above this value. 
Although the luminosity distributions overlap, the two populations are only 0.85\% likely to come from the same distribution, based on a K-S test (Figure \ref{lumHist}, inset). Ionizing radiation from the AGN increases the average luminosity of Ly$\alpha$ emission in our sample beyond what is typically observed in LAEs. However, the luminosity distribution overlap indicates that the covering factor is significant for the NLR in approximately half of the quasar host galaxies in our sample with narrow Ly$\alpha$ emission detected. 
This favors the idea that the majority of associated DLAs ($\sim 3/4$) are extended and probably due to galaxies neighboring the quasar host. Only the most luminous Ly$\alpha$ emission peaks, where the high luminosity implies a low covering factor for the NLR, support the idea of compact associated DLAs.

\subsection{Comparison with Radio Galaxies}

Radio galaxies are known to be surrounded by extended Ly$\alpha$ nebulosities \citep{2006AN....327..175V}. The AGN is hidden, and consequently most of the emission is from surrounding gas ionized by the central AGN. Some radio galaxies also exhibit Ly$\alpha$ absorption but with \ion{H}{I} column densities  lower than those in our sample \citep{1995MNRAS.277..389R}.
%In a unified model for AGN (ref) where what we observe depends on the viewing angle, radio galaxies are analogous to quasar host galaxies, except with a jet oriented in our direction. 
We therefore compare the Ly$\alpha$ emission detected in our sample to that of 61 $2.0 \leq z \leq 3.5$ radio galaxies from the \citet{2000A&A...362..519D} sample. 
%The luminosities from our sample are similar to those of radio galaxies with moderate luminosities, but the radio galaxy luminosities extend well beyond even the highest luminosity Ly$\alpha$ emission peak detected in our sample (Figure \ref{lumHist}). 
%Although the radio galaxy luminosities extend well beyond even the highest luminosity Ly$\alpha$ emission peak detected in our sample (Figure \ref{lumHist}), our luminosities coincide with the moderate luminosities for radio galaxies.
%However at moderate luminosities, the radio galaxy distribution coincides with the luminosities from our sample.
The luminosities from our sample coincide in part with the luminosity distribution for radio galaxies, overlapping at moderate luminosities. However, the radio galaxy luminosities extend well beyond even the highest luminosity Ly$\alpha$ emission peak detected in our sample (Figure \ref{lumHist}). It is not surprising that the radio galaxies have Ly$\alpha$ luminosities that surpass the luminosities we measure from narrow Ly$\alpha$ emission in quasar host galaxies, since the radio jet sends a huge amount of energy across several tens of kiloparsecs and provokes far-ranging Ly$\alpha$ emission. In quasar host galaxies, the AGN influence is predominantly at the heart of the galaxy in the NLR.

%Four of the 26 quasars in the associated DLA sample are radio-loud, based on a matching with FIRST. 

%
%______________________________________________________________

\section{DLAs with partial coverage} 

\begin{figure*}[!htbp]
\centering
\includegraphics[width=18cm]{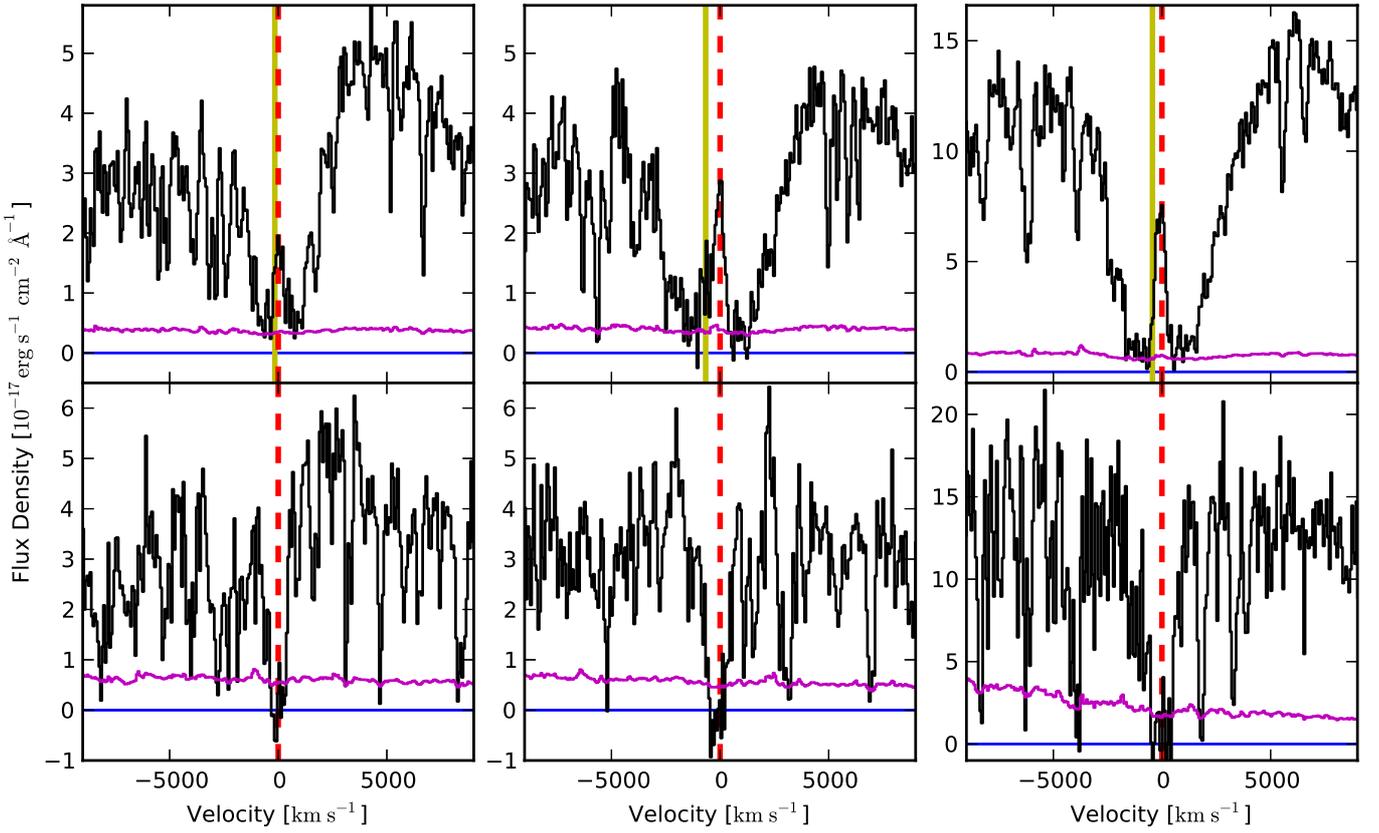}
\caption{Damped Ly$\alpha$ (top) and Ly$\beta$ (bottom) absorptions in the SDSS J0839+2709 (left), J1253+1007 (center), and J1323+2733 (right) spectra. In each case, the spectral flux density in the DLA trough is well-above the zero-level, while the Ly$\beta$ absorption is at zero. No zero-level offsets are applied here. All three DLAs reveal narrow Ly$\alpha$ emission. A dashed red line marks the associated DLA, and a solid yellow line indicates the velocity offset of the quasar.}
\label{nonZeroExamples}
\end{figure*}

\begin{figure*}[!htbp]
\centering
\includegraphics[width=18cm]{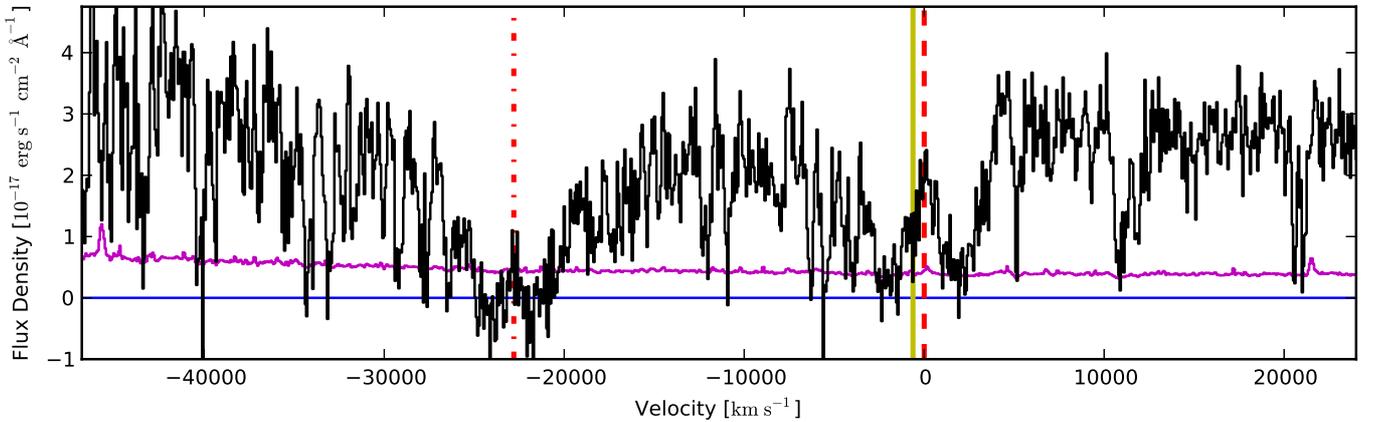}
\caption{Two consecutive DLAs in the SDSS J0148+1412 spectrum that both reveal narrow Ly$\alpha$ emission. A dashed (dash-dotted) red line marks the  associated (intervening) DLA, and a solid yellow line indicates the velocity offset of the quasar. The spectral flux density in the associated DLA trough is clearly above zero, whereas the trough of the intervening system at $\Delta~v~\sim~-22\,800~\text{km s}^{-1}$ goes to zero. No zero-level offset is applied in this Figure. }
\label{bestNonZero}
\end{figure*}

If the DLA does not cover the whole BLR entirely, then some residual flux (apart from the narrow Ly$\alpha$ emission) could be seen in the bottom of the Ly$\alpha$ trough. Estimating the exact value of the spectral flux density in the DLA trough is difficult, since the zero-level is not well-defined in SDSS spectra \citep{2011A&A...530A..50P, 2012A&A...548A..66P}.
 
In a few specific cases, however, we are able to demonstrate that some residual flux is indeed present and not due to data calibration errors. The covering factor appears to be less than one for several associated DLAs. The troughs of three DLAs with detected narrow Ly$\alpha$ emission are well-above zero at their bottoms, while their associated Ly$\beta$ absorptions are consistent with zero within errors (Figure \ref{nonZeroExamples}). The bottoms of the DLA troughs are flat, even though they are not at zero. Some BLR flux is probably transmitted, raising the spectral flux density level in the DLA and Ly$\beta$ absorptions. However, since the BLR Ly$\beta$ flux is more than five times weaker than the Ly$\alpha$ flux \citep{2001AJ....122..549V}, this effect on the Ly$\beta$ absorption is lost in the noise. If the non-zero DLA trough were due to quasar continuum emission, which follows a power law, we would expect the Ly$\beta$ absorption to also be clearly above zero. 
The observations indicate that the covering factor is less than one for these DLAs. 

%With a medium-resolution X-shooter spectrum, 
Alternatively, if the BLR is fully covered, the residual flux could be continuum emission from the quasar host galaxy. Indeed, \citet{2011A&A...532A..51Z} suggest that this possibility 
could account for the trough of the associated DLA towards Q0151+048A not going to zero.

%\citet{2011A&A...532A..51Z} suggest that the flux in the trough of an associated DLA towards Q0151+048A, which also does not go to zero, is continuum emission is from the quasar host galaxy.

%\citet{2011A&A...532A..51Z} also find that the flux in the trough of an associated DLA towards Q0151+048A does not go to zero. They identify a Ly$\alpha$ blob in the vicinity of the quasar as the source of Ly$\alpha$ emission in the DLA trough and suggest that residual continuum emission is from the quasar host galaxy.

\begin{figure}[!htbp]
%\resizebox{\hsize}{!}{\includegraphics{nonZeroAndInterDLA-noEm}}
\centering
\includegraphics{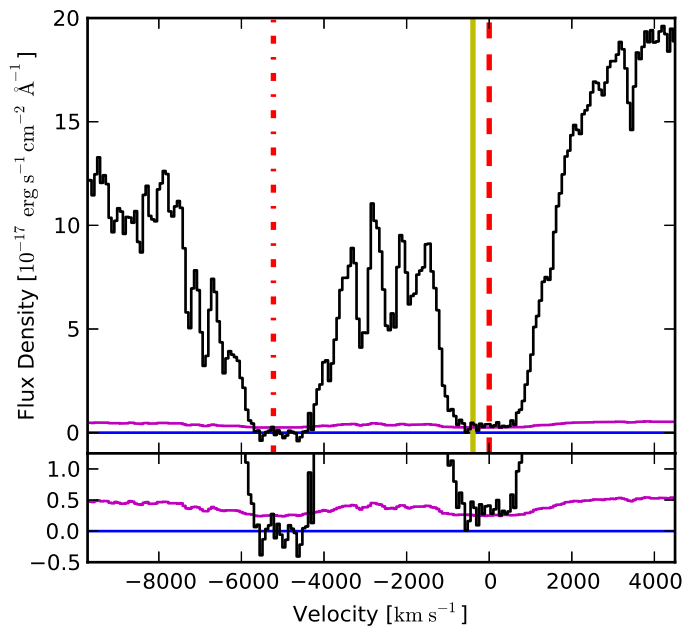}
\caption{Two consecutive DLAs in the SDSS J1635+1634 spectrum, neither of which have narrow Ly$\alpha$ emission detected. The spectral flux density in the associated DLA trough is slightly above zero, in contrast to the trough of the intervening DLA at $\Delta~v~\sim~-5025~\text{km s}^{-1}$ (see lower panel zoom). Labeling is the same as in Figure \ref{bestNonZero}.}
\label{neighborAtZero}
\end{figure}

In the most striking example towards SDSS~J0148+1412, an intervening DLA with a trough at zero occurs along the same line-of-sight as an associated DLA with a trough distinctly above zero (Figure \ref{bestNonZero}). Remarkably, both the intervening and the associated DLA show narrow Ly$\alpha$ emission. 
The line-of-sight towards SDSS~J1635+1634 (Figure \ref{neighborAtZero}) also has an associated and an intervening DLA, separated by only $\sim$5\,025 $\text{km s}^{-1}$. However, no narrow Ly$\alpha$ emission is detected in either DLA trough. Again, the bottom of the intervening DLA trough is at zero, while the associated DLA trough is slightly above the zero-level. If this offset in the associated DLA flux is due to partial coverage, we would expect to also detect narrow Ly$\alpha$ emission. Potentially, the covering factor is one and the flux from the lower-redshift galaxy creates the small zero-level offset in the associated DLA. Another explanation could be that the quasar has no detectable Ly$\alpha$ emission.

Partial coverage of absorption systems is an interesting phenomenon that can constrain the relative sizes of the foreground and background objects \citep[see e.g.,][]{2011MNRAS.418..357B}. The absorber radius, $r_{\rm DLA}$, along with the column density, can then be used to estimate the \ion{H}{I} number density, $n_{\ion{H}{I}}$, assuming a spherical absorber: 
$n_{\ion{H}{I}} = N({\ion{H}{I}}) / r_{\rm DLA}$.
%\newline 
%\noindent
The average \ion{H}{I} column density for the associated DLAs with detected narrow Ly$\alpha$ emission is $10^{21.53}\ \rm cm^{-2}$. If the DLA absorber is on the order of 10 - 100 pc, then $n_{\ion{H}{I}} \simeq 11 - 110\ \text{cm}^{-3}$. For smaller-sized absorbers, we could expect to find $\rm H_{2}$ gas, since it occurs in systems with $n_{\ion{H}{I}} \gtrsim 100 \ \text{cm}^{-3}$ \citep{2005MNRAS.362..549S}. However, due to the low resolution and S/N in SDSS spectra, only rare systems with $\rm H_{2}$ column density log \textit{N}$\rm (H_2) \gtrsim 18.5\ cm^{-2}$ can be detected (Balashev et al., in prep).

%As it will be difficult to achieve much beyond these detections with SDSS data, we intend to observe particularly interesting objects with X-shooter at the VLT \citep{2011A&A...536A.105V}.

\begin{figure*}
\centering
\includegraphics[width=18cm]{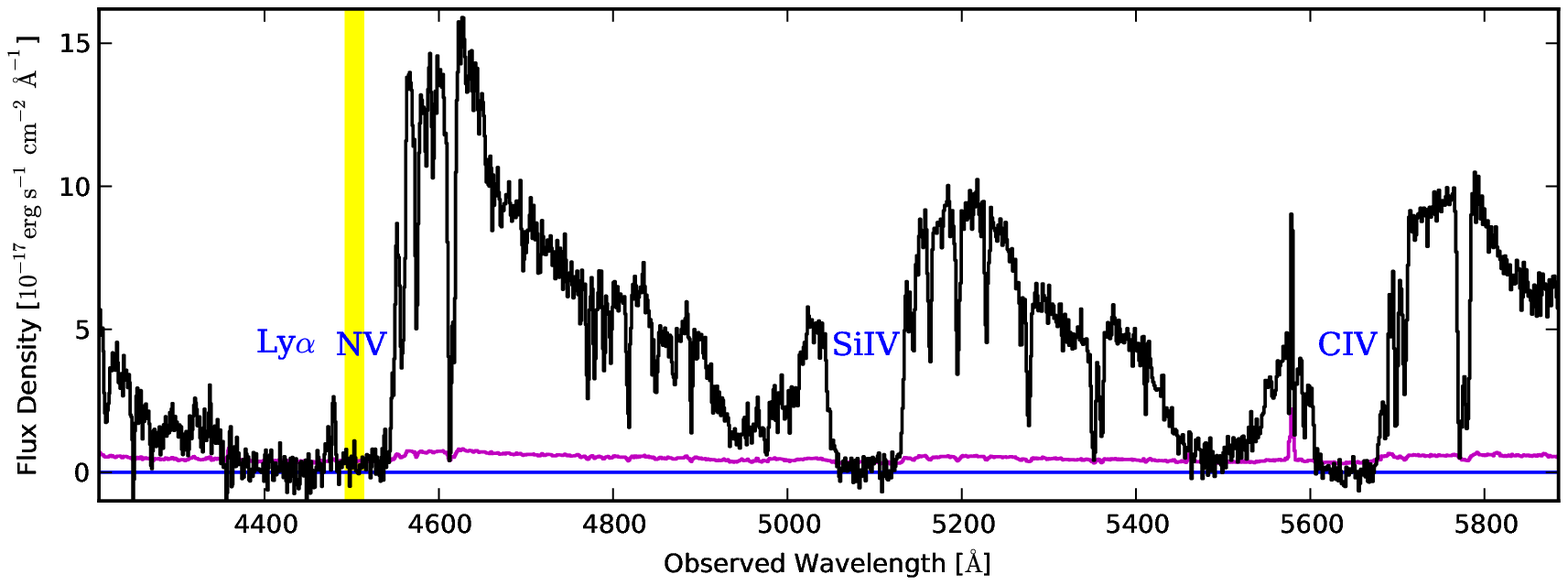}
\caption{BAL quasar SDSS J1116+3205 where no narrow Ly$\alpha$ emission is detected, despite absorption troughs that go to zero (labeled). Yellow shading indicates the possible range for the quasar Ly$\alpha$ emission peak, which coincides with the \ion{N}{V} BAL absorption trough.}
\label{noEmBAL}
\end{figure*}

%
%______________________________________________________________

\section{Discussion and Conclusions}

Thanks to the large number of z > 2 quasar lines-of-sight surveyed with SDSS-III BOSS, we have uncovered an unprecedented sample of strong associated DLAs that reveal narrow Ly$\alpha$ emission superimposed in their troughs. The number of associated DLAs in our statistical sample exceeds the anticipated number calculated from the $N(\ion{H}{I})$ distribution function by $\sim2.4$. This overdensity of DLAs is consistent with expectations based on what we know about galaxy clustering. 

We propose that when no narrow Ly$\alpha$ emission is detected in the trough of an associated DLA the absorber is a galaxy near to the quasar host, and when the absorber blocks the $\sim$1~pc BLR without also obscuring the extended Ly$\alpha$ emission the associated DLA arises from dense clouds of \ion{H}{I} gas in the quasar host galaxy. However, efforts to distinguish two populations of DLA sources from the statistical sample are inconclusive. DLAs with the largest $\text{W}_{0, \ion{Si}{II}}$ and $\text{W}_{0, \ion{Al}{II}}$ have no Ly$\alpha$ emission, but this trend does not hold for $\text{W}_{0, \ion{C}{II}}$ and $\text{W}_{0, \ion{Fe}{II}}$. The low resolution and S/N of BOSS spectra limits our ability to investigate this issue. With higher resolution and higher S/N spectra, we would be able to measure the column densities and study properties, such as the depletion, in greater detail.

Geometrical effects between the quasar host and a nearby galaxy could produce a configuration that transmits the narrow Ly$\alpha$ emission. Potentially, the weakest narrow Ly$\alpha$ emission occurs when the absorber is a nearby galaxy and the strongest narrow Ly$\alpha$ emission occurs when the absorber is an \ion{H}{I} cloud associated with the galactic environment. A continuum of such configurations would explain the lack of distinct characteristics for the two absorber populations.

%Another alternative could be that all associated DLAs arise from galaxies near to the quasar host, but in the cases with a detected narrow Ly$\alpha$ emission peak, the emitting region is significantly extended, beyond 10~kpc. We can investigate this possibility with long slit spectroscopy.

Additionally, the NLR may be significantly extended in quasar host galaxies, beyond 10~kpc \citep{2004ApJ...614..558N, 2013arXiv1307.5852H}. \citet{2011A&A...532A..51Z} determined that a Ly$\alpha$ blob located more than $\sim 30$~kpc from Q0151+048A is both the source of Ly$\alpha$ emission in the DLA trough and associated with the quasar host galaxy. With long slit spectroscopy, we could pinpoint the location of the narrow Ly$\alpha$ emission relative to the quasar and investigate the extent of emission regions in quasar host galaxies. 

The symmetrical profile of the narrow Ly$\alpha$ emission peaks, which is strikingly different from what is seen in LBGs, supports the idea that the gas in the DLA is not directly associated with the emission region and that the \ion{H}{I} content of this emission region is low. The line profile symmetry and lack of velocity offset for the Ly$\alpha$ emission peak with respect to the DLA center indicate that there is not a continuous distribution of \ion{H}{I} gas in the emission region to scatter the Ly$\alpha$ photons.

%the population in 
We have compared the Ly$\alpha$ luminosities in our sample to those of LAEs and radio galaxies. Roughly half of the narrow Ly$\alpha$ emission peaks detected by using DLAs as coronagraphs have luminosities consistent with standard LAEs. We argue that for these quasar host galaxies the DLA is extended enough to cover most of the NLR around the AGN and reduce the luminosity to a value more typical of a star-forming galaxy. The luminosities for the other half of our narrow Ly$\alpha$ emission peaks (25\% of the total sample) are higher than what is typical for LAEs. They are comparable to, but smaller than, those of radio galaxies. This is consistent with the fact that the AGN ionizes the surrounding gas, which in turn emits Ly$\alpha$ photons. In these cases, the DLA probably occurs in a small cloud in the host galaxy. Radio galaxies have higher luminosities, since their jets inject energy further from the center.

%\textbf{The similarity to the lower-end of radio galaxy luminosities indicates that the Ly$\alpha$ emission is extended.} \textit{(Dire plus? Mettre cette phrase ou?) }

Partial coverage in an associated DLA with no detected narrow Ly$\alpha$ emission raises the question of whether all quasars have a NLR, since a DLA absorber cannot leave the $\sim$1~pc BLR partially covered while simultaneously blocking emission across the entire NLR. We can potentially use BALs to investigate the possibility of quasars without emission from the NLR. Like associated DLAs, the gas clouds that create strong BALs in quasar spectra can serve as coronagraphs to obscure emission from the BLR and reveal NLR emission. The BAL clouds are located close to the central engine, so they cannot extend more than marginally into the NLR. 

\begin{figure}[b]
\centering
%\resizebox{\hsize}{!}{\includegraphics{velBAL_zBAL}}
\includegraphics{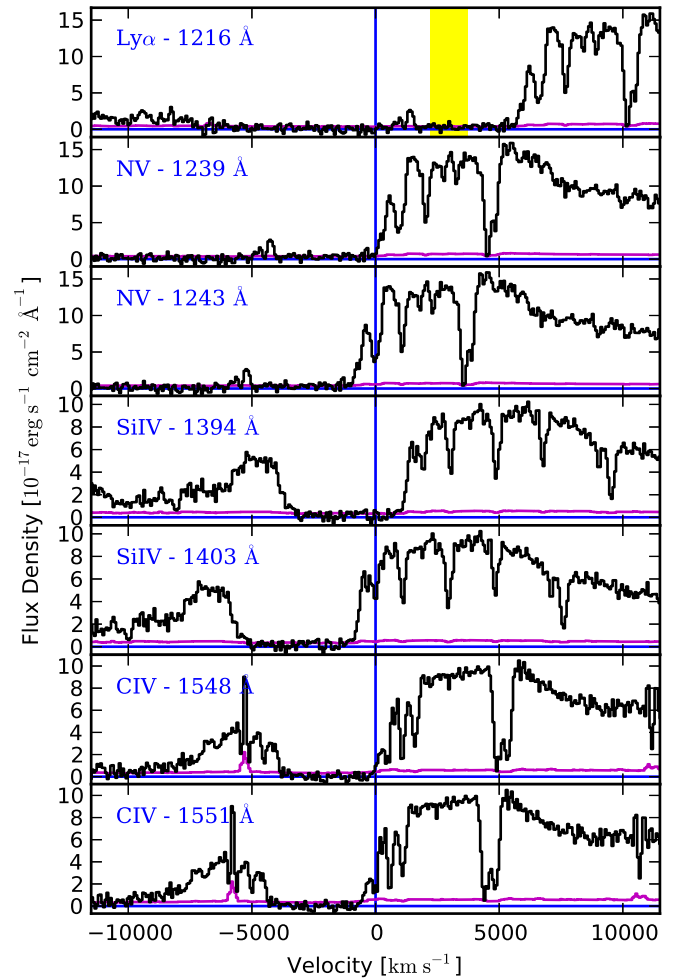}
\caption{Velocity plots for BAL quasar SDSS J1116+3205. The absorption system that indicates the beginning of the BAL troughs is at zero velocity. The possible redshift range for the quasar is shaded in yellow in the Ly$\alpha$ panel (top). The flux density is consistent with zero inside this region, which coincides with the \ion{N}{V} BAL trough. }
\label{velBALs}
\end{figure}

The difficulty with using BALs as coronagraphs, however, is that the spectral flux density in their \ion{H}{I} troughs seldom goes to zero, either because the optical depth is not high enough or because the cloud does not cover the BLR completely. We have searched for the very rare ideal occurrences where it is possible to measure Ly$\alpha$ flux from the NLR in a BAL trough. Figure \ref{noEmBAL} gives an example of a BAL quasar (SDSS~J111629.37+320511.5) where no emission is detected. We identify the minimum possible quasar redshift from the \ion{C}{III}] emission peak and limit the maximum to five times the typical blueshift for quasar \ion{C}{III}] emission peaks, 300 $\text{km s}^{-1}$ \citep{2012A&A...548A..66P}. The small rise in flux density at $\sim$4480~\AA\ corresponds to the transition between the \ion{N}{V} and Ly$\alpha$ BAL troughs (Figure \ref{velBALs}) and is unlikely to be narrow Ly$\alpha$ emission. This BAL quasar provides definite evidence that no Ly$\alpha$ emission is detected from the NLR of some quasars. 

Before further speculating about this intriguing fact, we must check if [\ion{O}{III}] emission is present in the BAL quasar spectra that lack narrow Ly$\alpha$ emission. X-shooter \citep{2011A&A...536A.105V} on the VLT would be the ideal instrument for obtaining additional spectroscopic data, since its wavelength coverage extends from the ultraviolet to the near-infrared.

Definitively interpreting why some associated DLAs reveal narrow Ly$\alpha$ emission while others do not is not possible with low-resolution spectra. Future observations with an Integral Field Unit (IFU) or higher resolution spectra taken with multiple position angles for each quasar are essential to make progress in this domain. Associated DLAs with narrow Ly$\alpha$ emission potentially offer a unique probe of the quasar host galaxy. The DLA provides a measure of the gas composition in the galactic environment, while emission from the NLR can be linked to ionization. Such systems could provide important insight into the effect of AGN feedback on their host galaxies.

%\textit{While it is likely that most associated DLAs arise from galaxies near to the quasar host, the minority that are potentially due to \ion{H}{I} clouds in the host galaxy environment could be of rich scientific value. Based on the present analysis, It is likely that most associated DLAs arise from galaxies near to the quasar host, a minority maybe be due to \ion{H}{I} clouds in the host galaxy environment. The present analysis also raises questions about the extent and strength of NLR emission in these quasar host galaxies. The narrow Ly$\alpha$ emission properties indicate that certain quasar host galaxies could have significantly extended narrow emission regions, while others may have surprisingly little NLR emission. Certain associated DLAs reveal narrow Ly$\alpha$ emission that indicates the quasar host galaxies could have significantly extended narrow emission regions, while others favor hosts with... }

\begin{acknowledgements}
We thank the anonymous referee for comments that helped to improve the article.

The SDSS-III French Participation Group was supported by the Agence Nationale de la Recherche under grants ANR-08-BLAN-0222 and ANR-12-BS05-0015.

Funding for SDSS-III has been provided by the Alfred P. Sloan Foundation, the Participating Institutions, the National Science Foundation, and the U.S. Department of Energy Office of Science. The SDSS-III web site is http://www.sdss3.org/.

SDSS-III is managed by the Astrophysical Research Consortium for the Participating Institutions of the SDSS-III Collaboration including the University of Arizona, the Brazilian Participation Group, Brookhaven National Laboratory, University of Cambridge, Carnegie Mellon University, University of Florida, the French Participation Group, the German Participation Group, Harvard University, the Instituto de Astrofisica de Canarias, the Michigan State/Notre Dame/JINA Participation Group, Johns Hopkins University, Lawrence Berkeley National Laboratory, Max Planck Institute for Astrophysics, Max Planck Institute for Extraterrestrial Physics, New Mexico State University, New York University, Ohio State University, Pennsylvania State University, University of Portsmouth, Princeton University, the Spanish Participation Group, University of Tokyo, University of Utah, Vanderbilt University, University of Virginia, University of Washington, and Yale University. 
\end{acknowledgements}

% Tables
%________________________________________________________________
\begin{longtab}
\centering 							% used for centering table
\begin{longtable}{l l l r l l c r}	% centered columns (4 columns)
\caption{\label{DLAtable} Associated DLAs}\\		% title of Table
\hline\hline 						% inserts double horizontal lines
SDSS Name\tablefootmark{a} & $z_{\rm \, QSO}$ & $\text{M}_{i}$ & E(B-V) & log $N(\ion{H}{I})$ & $z_{\rm \, DLA}$ & $z_{\rm \, Ly\alpha}$ & Integ. Flux\tablefootmark{b} \\ % table heading
          &               &                &        & ($\rm cm^{-2}$)      &               &  &  $\left(\rm \dfrac{10^{-17}\ erg}{s\ cm^{2}}\right)$ \\
\hline
\endfirsthead
\caption{continued.}\\
\hline\hline
SDSS Name\tablefootmark{a} & $z_{\rm \, QSO}$ & $\text{M}_{i}$ & E(B-V) & log $N(\ion{H}{I})$ & $z_{\rm \, DLA}$ & $z_{\rm \, Ly\alpha}$ & Integ. Flux\tablefootmark{b} \\ % table heading
          &               &                &        & ($\rm cm^{-2}$)      &               &  & $\left(\rm \dfrac{10^{-17}\ erg}{s\ cm^{2}}\right)$ \\
\hline
\endhead
\hline
\hline
%\endfoot
\noalign{\smallskip}	
\multicolumn{8}{c}{DR9 Associated DLAs} \\
\noalign{\smallskip}	
\hline
\noalign{\smallskip}	
\multicolumn{8}{c}{Statistical Sample} \\
\noalign{\smallskip}	
\hline								% inserts single horizontal line
013654.00-005317.5 & 2.491 & -26.43 & -0.076 & 21.30 & 2.493 &  ...  &  $\leq$ 4.6\\ 
014858.31+141235.4\tablefootmark{c} & 3.172 & -27.06 & -0.022 & 21.56 & 3.181 & 3.182 &  45.2\\ 
073446.02+322205.3 & 2.314 & -26.51 &  0.029 & 21.35 & 2.308 &  ...  &  $\leq$ 6.5\\ 
080523.32+214921.1 & 3.487 & -28.08 &  0.032 & 21.55 & 3.477 &  ...  &  $\leq$ 4.3\\ 
081626.02+351055.2 & 3.073 & -26.68 &  0.033 & 21.31 & 3.071 &  ...  &  $\leq$ 3.4\\ 
083716.51+195025.9 & 2.262 & -24.70 & -0.036 & 21.31 & 2.271 &  ...  &  $\leq$ 5.2\\ 
084448.59+043504.7 & 2.367 & -26.31 & -0.058 & 21.33 & 2.379 &  ...  &  $\leq$ 7.5\\ 
093859.15+405257.8 & 2.172 & -25.88 & -0.015 & 21.43 & 2.176 & 2.170 &  41.1\\ 
093924.71+430246.8 & 2.722 & -25.71 &  0.014 & 21.45 & 2.717 & 2.722 &  14.5\\ 
095307.13+034933.8 & 2.599 & -26.00 &  0.025 & 22.10 & 2.592 & 2.594 &  14.1\\ 
104939.76+433533.6 & 2.505 & -25.01 & -0.022 & 21.40 & 2.514 &  ...  &  $\leq$ 5.1\\ 
105138.65+433921.5 & 2.670 & -26.10 &  0.002 & 22.07 & 2.665 &  ...  &  $\leq$ 5.3\\ 
115605.93+354538.5 & 2.125 & -24.61 & -0.042 & 21.40 & 2.126 & 2.128 &  17.5\\ 
121902.52+382822.1 & 3.012 & -26.65 &  0.083 & 21.34 & 3.012 &  ...  &  $\leq$ 4.4\\ 
125635.73+350604.8 & 2.237 & -25.47 &  0.008 & 22.09 & 2.238 & 2.239 &  51.0\\ 
131309.25+392456.9 & 2.446 & -25.96 &  0.043 & 21.42 & 2.447 &  ...  &  $\leq$ 5.5\\ 
132922.21+052014.3 & 2.999 & -26.36 & -0.006 & 21.57 & 2.994 &  ...  &  $\leq$ 4.2\\ 
133430.83+383819.1 & 2.176 & -25.80 & -0.028 & 21.62 & 2.181 & 2.180 &  11.6\\ 
141745.90+362127.4 & 3.338 & -25.66 &  0.056 & 22.34 & 3.338 &  ...  &  $\leq$ 3.0\\ 
142542.06+025701.4 & 2.199 & -25.55 & -0.046 & 21.35 & 2.205 &  ...  &  $\leq$ 6.0\\ 
143253.28-005837.6 & 2.530 & -26.50 & -0.042 & 21.36 & 2.537 &  ...  &  $\leq$ 5.5\\ 
143634.65+350553.8 & 2.212 & -25.56 & -0.032 & 21.42 & 2.220 &  ...  &  $\leq$ 7.1\\ 
150016.39+070609.1 & 2.367 & -26.08 &  0.024 & 21.33 & 2.379 &  ...  &  $\leq$ 6.9\\ 
150049.63+053320.0 & 2.392 & -26.76 &  0.085 & 21.36 & 2.398 &  ...  &  $\leq$ 7.3\\ 
150227.84+054853.8 & 2.753 & -26.69 &  0.037 & 21.40 & 2.738 &  ...  &  $\leq$ 4.6\\ 
150812.79+363530.3 & 2.097 & -25.50 & -0.046 & 21.33 & 2.106 &  ...  &  $\leq$ 16.1\\ 
151129.27+181138.7 & 3.120 & -25.81 & -0.020 & 21.39 & 3.134 &  ...  &  $\leq$ 3.9\\ 
151843.30+050808.3 & 2.436 & -26.79 & -0.025 & 21.38 & 2.441 &  ...  &  $\leq$ 6.5\\ 
154621.59+020102.9 & 2.821 & -28.60 &  0.002 & 21.34 & 2.810 &  ...  &  $\leq$ 3.3\\ 
163217.98+115524.3 & 3.101 & -26.25 & -0.049 & 21.51 & 3.092 &  ...  &  $\leq$ 3.3\\ 
223037.78-000531.5 & 2.407 & -26.25 &  0.039 & 21.45 & 2.412 & 2.406 &   8.1\\ 
\noalign{\smallskip}	
\hline 
\noalign{\smallskip}	
020555.00+061849.3 & 2.223 & -25.09 & -0.084 & 20.91 & 2.223 &  ...  &  $\leq$ 7.5\\ 
075556.95+365658.3 & 2.935 & -25.65 & -0.081 & 21.25 & 2.973\tablefootmark{d} &  ...  &  $\leq$ 2.5\\ 
080045.73+341128.9 & 2.334 & -26.13 & -0.001 & 21.24 & 2.329 & 2.328 &  16.5\\ 
083523.26+141930.8 & 3.139 & -28.01 & -0.017 & 21.27 & 3.142 &  ...  &  $\leq$ 3.8\\ 
083954.59+270913.1 & 2.959 & -26.72 & -0.064 & 21.10 & 2.961 & 2.961 &   8.5\\ 
085600.89+030218.7 & 3.039 & -25.55 & -0.011 & 21.23 & 3.056 &  ...  &  $\leq$ 2.9\\ 
100112.40-014737.9 & 2.188 & -25.32 & -0.038 & 21.21 & 2.199 &  ...  &  $\leq$ 7.0\\ 
114639.00-010703.2 & 3.072 & -25.23 & -0.084 & 21.28 & 3.063 &  ...  &  $\leq$ 2.9\\ 
115656.97+024619.8 & 2.377 & -25.11 & -0.017 & 21.28 & 2.383 &  ...  &  $\leq$ 8.3\\ 
121716.25+012139.1 & 3.388 & -26.34 & -0.033 & 21.13 & 3.370 &  ...  &  $\leq$ 4.0\\ 
135138.57+352338.8 & 2.078 & -25.97 &  0.035\tablefootmark{e} & 21.26 & 2.074 &  ...  &  $\leq$ 10.4\\ 
145155.59+295311.7 & 2.185 & -25.50 &  0.013 & 21.18 & 2.169 &  ...  &  $\leq$ 9.0\\ 
150806.90+283835.4 & 2.341 & -25.39 & -0.023 & 21.15 & 2.341 &  ...  &  $\leq$ 7.1\\ 
152243.24+065138.2 & 2.146 & -26.46 &  0.018 & 21.25 & 2.148 &  ...  &  $\leq$ 11.3\\ 
155310.67+233013.3 & 3.022 & -27.39 & -0.001 & 21.29 & 3.023 & 3.025 &  10.7\\ 
221811.59+155420.1 & 2.294 & -25.59 & -0.023 & 20.79 & 2.302 & 2.304 &  25.4\\ 
222007.76+002331.8 & 2.432 & -25.74 &  0.108 & 20.97 & 2.434 &  ...  &  $\leq$ 4.8\\ 
225307.84+030843.1 & 2.327 & -25.51 &  0.005 & 21.01 & 2.320 &  ...  &  $\leq$ 3.9\\ 
234757.96+034910.7 & 2.182 & -26.38 & -0.072 & 21.52 & 2.155 &  ...  &  $\leq$ 9.0\\ 
\hline		
\hline 
\noalign{\smallskip}	
\multicolumn{8}{c}{Additional Associated DLAs with Ly$\alpha$ Emission Detected} \\
\noalign{\smallskip}
\hline							
005917.64+112407.7 & 3.030 & -26.66 &  0.015 & 21.65 & 3.034 & 3.039 &  10.4\\ 
011226.76-004855.8 & 2.137 & -25.25 & -0.016 & 21.82 & 2.150 & 2.148 &  64.7\\ 
082303.22+052907.6 & 3.188 & -25.97 &  0.090 & 21.61 & 3.191 & 3.193 &  66.4\\ 
083300.07+233538.5 & 2.459 & -26.17 &  0.116 & 21.95 & 2.460 & 2.450 &  50.7\\ 
091137.46+225821.4 & 2.185 & -25.82 &  0.000 & 21.51 & 2.188 & 2.190 &   9.0\\ 
103124.70+154118.1 & 2.151 & -25.77 &  0.038 & 21.44 & 2.155 & 2.156 & 141.7\\ 
105823.73+031524.4 & 2.301 & -25.00 & -0.068 & 21.79 & 2.293 & 2.305 &  27.1\\ 
112829.38+151826.5 & 2.246 & -26.43 &  0.076 & 21.59 & 2.251 & 2.245 &  56.9\\ 
115432.67-021538.0 & 2.187 & -24.94 & -0.078 & 21.79 & 2.186 & 2.184 &  46.2\\ 
125302.00+100742.3 & 3.023 & -27.04 & -0.000 & 21.24 & 3.032 & 3.031 &  23.5\\ 
132303.48+273301.9 & 2.625 & -27.46 & -0.003 & 21.36 & 2.631 & 2.630 &  49.4\\ 
150217.33+084715.4 & 2.264 & -25.40 &  0.084 & 21.21 & 2.264 & 2.267 &   9.3\\ 
225918.46-014534.8 & 2.508 & -26.10 & -0.030 & 21.62 & 2.502 & 2.509 & 215.9\\ 
230709.81+020843.0 & 2.157 & -25.00 &  0.000 & 21.80 & 2.165 & 2.164 &  11.9\\ 
\hline
\noalign{\smallskip}	
\multicolumn{8}{c}{Additional Associated DLA with No Detected Ly$\alpha$ Emission (Figure \ref{neighborAtZero})} \\
\noalign{\smallskip}
\hline		
163538.57+163436.9\tablefootmark{c} & 3.019 & -28.63 & -0.005 & 20.91 & 3.024 &  ...  &  $\leq$ 3.4\\ 
\hline						
\end{longtable}
\tablefoot{
\tablefoottext{a}{J2000 coordinates}
\tablefoottext{b}{4-$\sigma$ upper limit on the integrated flux when no Ly$\alpha$ emission is detected.}
\tablefoottext{c}{Additional intervening DLA along the line-of-sight}
\tablefoottext{d}{$z_{\,\rm DLA}$ fitted with $N(\ion{H}{I})$, not from metals}
\tablefoottext{e}{An LMC, rather than SMC, reddening law may be preferred in this case.}}
\end{longtab}

%
%________________________________________________________________

\begin{longtab}
\begin{landscape}
\begin{longtable}{l l l l l l l l l l l l l l}		\\
\caption{\label{W0table} Rest Equivalent Widths\tablefootmark{a}}  	\\		% title of Table		
\hline\hline 						% inserts double horizontal lines
SDSS Name & $z_{\rm \, DLA}$\tablefootmark{b} & \multicolumn{3}{c}{\ion{C}{II} -- 1334$\AA$} & \multicolumn{3}{c}{\ion{Si}{II} -- 1526$\AA$} & \multicolumn{3}{c}{\ion{Al}{II} -- 1670$\AA$} & \multicolumn{3}{c}{\ion{Fe}{II} -- 2344$\AA$}\\
  &   & Fit ($\AA$) & $\mathcal{F}$ ($\AA$) & Err ($\AA$) & Fit ($\AA$) & \spef\ ($\AA$) & Err ($\AA$) & Fit ($\AA$) & \spef\ ($\AA$) & Err ($\AA$) & Fit ($\AA$) & \spef\ ($\AA$) & Err ($\AA$)\\
%  &   & $\AA$ & $\AA$ & $\AA$ & $\AA$ & $\AA$ & $\AA$ & $\AA$ & $\AA$ & $\AA$ & $\AA$ & $\AA$ & $\AA$ \\
\hline
\endfirsthead
\caption{continued.}\\
\hline\hline
SDSS Name & $z_{\rm \, DLA}$\tablefootmark{b} & \multicolumn{3}{c}{\ion{C}{II} -- 1334$\AA$} & \multicolumn{3}{c}{\ion{Si}{II} -- 1526$\AA$} & \multicolumn{3}{c}{\ion{Al}{II} -- 1670$\AA$} & \multicolumn{3}{c}{\ion{Fe}{II} -- 2344$\AA$}\\
  &   & Fit ($\AA$) & \spef\ ($\AA$) & Err ($\AA$) & Fit ($\AA$) & \spef\ ($\AA$) & Err ($\AA$) & Fit ($\AA$) & \spef\ ($\AA$) & Err ($\AA$) & Fit ($\AA$) & \spef\ ($\AA$) & Err ($\AA$)\\
%  &   &  Fit  &  Flux & Err   & Fit   & Flux  & Err   & Fit   & Flux  & Err   & Fit   & Flux  & Err \\
%  &   & $\AA$ & $\AA$ & $\AA$ & $\AA$ & $\AA$ & $\AA$ & $\AA$ & $\AA$ & $\AA$ & $\AA$ & $\AA$ & $\AA$ \\
\hline
\endhead
\hline
\noalign{\smallskip}	
\multicolumn{14}{c}{DR9 Associated DLAs} \\
\noalign{\smallskip}	
\hline
\noalign{\smallskip}	
\multicolumn{14}{c}{Statistical Sample} \\
\noalign{\smallskip}	
\hline									% inserts single horizontal line
013654.00-005317.5 & 2.493 & 2.198 & 2.252 & 0.103 & 1.882 & 1.992 & 0.084 & 2.040 & 2.051 & 0.161 & 2.625 & 2.600 & 0.258\\ 
014858.31+141235.4 & 3.181 &   ... &   ... &   ... & 1.384 & 1.224 & 0.089 & 1.379 & 1.318 & 0.107 & 2.323 & 3.020 & 0.246\\ 
073446.02+322205.3 & 2.308 & 2.292 & 2.106 & 0.099 & 1.759 & 1.473 & 0.080 & 2.194 & 2.665 & 0.094 & 2.229 & 2.014 & 0.190\\ 
080523.32+214921.1 & 3.477 & 1.511 & 1.560 & 0.081 & 1.163 & 0.984 & 0.069 & 0.906 & 1.037 & 0.091 &   ... &   ... &   ...\\ 
081626.02+351055.2 & 3.071 & 0.494 & 0.760 & 0.114 & 0.534 & 0.565 & 0.112 & 0.262 & 0.324 & 0.118 & 0.192 & 0.365 & 0.452\\ 
083716.51+195025.9 & 2.271 & 0.997 & 1.044 & 0.327 & 0.799 & 0.902 & 0.195 & 0.854 & 1.241 & 0.549 & 2.477 & 1.916 & 1.115\\ 
084448.59+043504.7 & 2.379 & 0.249 & 0.161 & 0.099 & 0.378 & 0.436 & 0.087 & 0.380 & 0.337 & 0.122 & 0.881 & 1.368 & 0.251\\ 
093859.15+405257.8 & 2.176 & 0.546 & 0.576 & 0.109 & 0.437 & 0.452 & 0.084 & 0.563 & 0.542 & 0.111 & 0.641 & 0.676 & 0.180\\ 
093924.71+430246.8 & 2.717 & 0.713 & 0.777 & 0.240 & 0.765 & 0.826 & 0.215 & 0.132 & 0.208 & 0.316 & 0.242 & 0.605 & 0.688\\ 
095307.13+034933.8 & 2.592 & 1.254 & 1.520 & 0.157 & 1.457 & 1.419 & 0.124 & 1.387 & 1.478 & 0.223 & 1.359 & 0.726 & 0.340\\ 
104939.76+433533.6 & 2.514 & 0.556 & 0.878 & 0.272 & 0.407 & 0.288 & 0.169 & 0.412 & 0.710 & 0.496 & 0.786 & 2.066 & 0.612\\ 
105138.65+433921.5 & 2.665 & 0.586 & 0.903 & 0.134 & 0.736 & 1.115 & 0.109 & 0.480 & 0.828 & 0.225 & 1.453 & 1.433 & 0.272\\ 
115605.93+354538.5 & 2.126 & 2.530 & 2.561 & 0.118 & 1.274 & 1.269 & 0.089 & 1.537 & 1.626 & 0.127 & 1.215 & 1.865 & 0.232\\ 
121902.52+382822.1 & 3.012 &   ... & 2.783 & 0.111 & 1.930 & 2.018 & 0.143 & 1.927 & 1.843 & 0.150 & 2.279 & 1.977 & 0.694\\ 
125635.73+350604.8 & 2.238 &   ... & 1.181 & 0.119 & 0.774 & 0.737 & 0.096 & 0.741 & 0.472 & 0.120 & 1.175 & 1.174 & 0.226\\ 
131309.25+392456.9 & 2.447 & 1.896 & 1.944 & 0.190 & 2.663 & 2.727 & 0.116 & 1.551 & 1.580 & 0.191 & 1.411 & 1.233 & 0.376\\ 
132922.21+052014.3 & 2.994 & 2.184 & 2.048 & 0.132 & 1.881 & 1.930 & 0.130 & 1.824 & 1.797 & 0.137 & 2.385 & 3.842 & 1.271\\ 
133430.83+383819.1 & 2.181 & 0.392 & 0.454 & 0.077 & 0.271 & 0.329 & 0.067 & 0.184 & 0.267 & 0.084 & 0.300 & 0.321 & 0.138\\ 
141745.90+362127.4 & 3.338 & 1.413 & 1.478 & 0.287 & 1.289 & 1.334 & 0.185 & 1.592 & 1.803 & 0.484 &   ... &   ... &   ...\\ 
142542.06+025701.4 & 2.205 & 0.908 & 0.954 & 0.147 & 1.141 & 1.233 & 0.096 & 0.687 & 0.701 & 0.171 & 1.711 & 1.932 & 0.468\\ 
143253.28-005837.6 & 2.537 & 1.098 & 1.244 & 0.078 & 0.884 & 0.726 & 0.064 & 0.701 & 0.652 & 0.113 & 1.351 & 1.139 & 0.179\\ 
143634.65+350553.8 & 2.220 & 0.459 & 0.594 & 0.130 & 0.599 & 0.724 & 0.086 & 0.539 & 0.508 & 0.116 & 0.557 & 0.568 & 0.245\\ 
150016.39+070609.1 & 2.379 & 2.572 & 2.429 & 0.169 & 1.774 & 1.702 & 0.121 & 1.951 & 2.077 & 0.184 & 1.285 & 3.312 & 0.541\\ 
150049.63+053320.0 & 2.398 & 2.192 & 2.635 & 0.114 & 1.864 & 1.912 & 0.062 & 1.990 & 2.080 & 0.093 & 2.273 & 2.801 & 0.166\\ 
150227.84+054853.8 & 2.738 & 0.693 & 0.705 & 0.111 & 0.786 & 0.860 & 0.101 & 0.616 & 0.533 & 0.141 & 1.046 & -0.150 & 0.168\\ 
150812.79+363530.3 & 2.106 & 0.357 & 0.460 & 0.120 & 0.243 & 0.181 & 0.087 & 0.168 & 0.155 & 0.097 & 0.537 & 0.300 & 0.191\\ 
151129.27+181138.7 & 3.134 & 0.643 & 0.649 & 0.229 & 0.495 & 0.288 & 0.193 & 0.669 & 0.621 & 0.285 & 0.156 & 1.131 & 0.631\\ 
151843.30+050808.3 & 2.441 & 0.903 & 1.009 & 0.062 & 0.958 & 1.014 & 0.087 & 0.855 & 0.895 & 0.065 & 1.371 & 1.473 & 0.138\\ 
154621.59+020102.9 & 2.810 & 0.648 & 0.720 & 0.035 & 0.654 & 0.644 & 0.038 & 0.743 & 0.809 & 0.050 & 1.077 & 0.977 & 0.085\\ 
163217.98+115524.3 & 3.092 & 1.319 & 2.150 & 0.249 & 1.219 & 1.969 & 0.168 & 0.398 & 2.005 & 0.268 & 1.286 & -0.153 & 1.541\\ 
223037.78-000531.5 & 2.412 & 0.896 & 1.117 & 0.141 & 0.680 & 0.589 & 0.105 & 0.772 & 0.558 & 0.160 & 0.747 & 0.479 & 0.367\\ 
\noalign{\smallskip}	
\hline 
\noalign{\smallskip}
221811.59+155420.1 & 2.302 &   ... & 0.937 & 0.160 & 0.632 & 0.644 & 0.098 & 0.283 & 0.027 & 0.165 & 0.559 & 0.502 & 0.364\\ 
020555.00+061849.3 & 2.223 & 0.241 & 0.505 & 0.188 & 0.413 & 0.593 & 0.166 & 0.219 & 0.414 & 0.291 & 1.385 & 1.534 & 1.019\\ 
222007.76+002331.8 & 2.434 & 2.719 & 3.338 & 0.180 & 1.804 & 2.068 & 0.111 & 1.953 & 2.030 & 0.183 & 1.968 & 2.753 & 0.205\\ 
225307.84+030843.1 & 2.320 & 1.069 & 0.987 & 0.195 & 1.205 & 1.233 & 0.116 & 1.477 & 1.310 & 0.212 & 1.906 & 3.069 & 0.483\\ 
083954.59+270913.1 & 2.961 & 0.455 & 0.618 & 0.119 &   ... &   ... &   ... & 0.242 & 0.373 & 0.162 & 0.848 & 0.710 & 0.375\\ 
121716.25+012139.1 & 3.370 & 1.694 & 1.651 & 0.254 & 1.794 & 1.926 & 0.204 & 1.552 & 1.387 & 0.324 &   ... &   ... &   ...\\ 
150806.90+283835.4 & 2.341 & 0.410 & 0.382 & 0.101 & 0.373 & 0.371 & 0.074 & 0.449 & 0.286 & 0.424 & 0.417 & 0.408 & 0.208\\ 
145155.59+295311.7 & 2.169 & 0.391 & 0.677 & 0.193 & 0.286 & 0.327 & 0.133 & 0.250 & 0.263 & 0.173 & 0.219 & 0.427 & 0.385\\ 
100112.40-014737.9 & 2.199 & 0.943 & 1.224 & 0.145 & 1.071 & 1.149 & 0.097 & 1.313 & 1.235 & 0.139 & 1.267 & 1.358 & 0.269\\ 
085600.89+030218.7 & 3.056 & 3.607 & 3.857 & 0.212 & 3.038 & 4.849 & 0.211 & 2.981 & 2.668 & 0.276 &   ... &   ... &   ...\\ 
080045.73+341128.9 & 2.329 & 0.945 & 1.096 & 0.110 & 0.781 & 0.719 & 0.074 & 0.827 & 0.900 & 0.121 & 0.808 & 0.528 & 0.199\\ 
075556.95+365658.3\tablefootmark{c} &   ... &   ... &   ... &   ... &   ... &   ... &   ... &   ... &   ... &   ... &   ... &   ... &   ...\\ 
152243.24+065138.2 & 2.148 & 2.031 & 2.748 & 0.103 & 1.211 & 1.071 & 0.061 & 1.159 & 1.210 & 0.082 & 1.412 & 1.397 & 0.114\\ 
135138.57+352338.8 & 2.074 & 1.517 & 1.567 & 0.114 & 1.368 & 1.428 & 0.068 & 1.514 & 1.479 & 0.081 & 2.150 & 2.190 & 0.135\\ 
083523.26+141930.8 & 3.142 & 0.725 & 0.778 & 0.045 & 0.701 & 0.685 & 0.041 & 0.710 & 0.675 & 0.054 & 1.050 & 0.565 & 0.197\\ 
114639.00-010703.2 & 3.063 & 0.893 & 1.108 & 0.283 & 0.467 & 1.328 & 0.379 & 0.414 & 0.805 & 0.420 &   ... &   ... &   ...\\ 
115656.97+024619.8 & 2.383 & 1.303 & 1.551 & 0.180 & 1.052 & 0.983 & 0.111 & 1.357 & 1.401 & 0.197 & 0.513 & 1.172 & 0.346\\ 
155310.67+233013.3 & 3.023 & 0.205 & 0.191 & 0.084 & 0.188 & 0.239 & 0.079 & 0.287 & 0.286 & 0.080 & 0.331 & 0.378 & 0.212\\ 
234757.96+034910.7 & 2.155 & 1.583 & 1.813 & 0.082 & 1.395 & 1.336 & 0.072 & 1.307 & 1.305 & 0.076 & 2.042 & 1.825 & 0.238\\ 
\hline 
\noalign{\smallskip}	
\multicolumn{14}{c}{Additional Associated DLAs with Ly$\alpha$ Emission Detected} \\
\noalign{\smallskip}
\hline 					
005917.64+112407.7 & 3.034 & 1.014 & 1.375 & 0.129 & 0.803 & 0.656 & 0.158 & 0.820 & 0.786 & 0.127 & 1.381 & 1.490 & 0.554\\ 
011226.76-004855.8 & 2.150 & 0.626 & 1.116 & 0.294 & 0.714 & 0.906 & 0.180 & 0.667 & 0.820 & 0.289 & 0.415 & -0.096 & 0.614\\ 
082303.22+052907.6 & 3.191 & 2.666 & 2.809 & 0.328 & 1.569 & 1.683 & 0.197 & 2.479 & 2.108 & 0.281 &   ... &   ... &   ...\\ 
083300.07+233538.5 & 2.460 & 1.066 & 1.350 & 0.206 &   ... &   ... &   ... & 1.239 & 1.251 & 0.201 & 1.501 & 1.349 & 0.199\\ 
091137.46+225821.4 & 2.188 & 1.853 & 1.806 & 0.119 & 1.307 & 1.367 & 0.081 & 1.007 & 0.918 & 0.133 & 0.791 & 1.057 & 0.315\\ 
103124.70+154118.1 & 2.155 & 1.342 & 2.089 & 0.158 & 1.022 & 0.922 & 0.104 & 1.115 & 1.189 & 0.120 & 1.124 & 1.388 & 0.247\\ 
105823.73+031524.4 & 2.293 & 0.580 & 1.138 & 0.279 & 0.420 & 0.771 & 0.245 &   ... &   ... &   ... & 0.887 & 3.283 & 0.776\\ 
112829.38+151826.5 & 2.251 & 3.052 & 3.148 & 0.096 & 2.547 & 2.382 & 0.067 & 2.353 & 2.379 & 0.087 & 2.631 & 2.325 & 0.288\\ 
115432.67-021538.0 & 2.186 &   ... & 2.186 & 0.248 & 0.200 & 0.672 & 0.309 & 0.563 & 0.661 & 0.350 & 0.074 & -0.163 & 1.063\\ 
125302.00+100742.3 & 3.032 & 0.992 & 1.376 & 0.124 & 0.723 & 0.587 & 0.105 & 0.741 & 0.718 & 0.122 & 0.804 & 0.905 & 0.283\\ 
132303.48+273301.9 & 2.631 & 0.284 & 0.462 & 0.044 & 0.310 & 0.364 & 0.040 & 0.381 & 0.312 & 0.061 & 0.494 & 0.377 & 0.081\\ 
150217.33+084715.4 & 2.264 & 0.822 & 1.475 & 0.238 & 0.943 & 0.933 & 0.161 & 0.867 & 0.969 & 0.188 & 1.400 & 0.696 & 0.397\\ 
225918.46-014534.8 & 2.502 &   ... & 2.388 & 0.106 & 1.959 & 1.788 & 0.089 & 1.996 & 1.872 & 0.134 & 2.421 & 1.963 & 0.465\\ 
230709.81+020843.0 & 2.165 & 1.340 & 2.085 & 0.358 & 0.980 & 1.437 & 0.423 & 4.088 & 4.363 & 0.310 & 1.623 & 1.612 & 0.967\\ 
\hline
\noalign{\smallskip}	
\multicolumn{14}{c}{Additional Associated DLA with No Detected Ly$\alpha$ Emission (Figure \ref{neighborAtZero})} \\
\noalign{\smallskip}
\hline		
\noalign{\smallskip}	
163538.57+163436.9 & 3.024 & 0.270 & 0.255 & 0.030 & 0.215 & 0.182 & 0.029 & 0.225 & 0.200 & 0.031 & 0.424 & 0.781 & 0.082\\ 
\hline	
\end{longtable}
\tablefoot{
\tablefoottext{a}{Measured from the Voigt profile fit and from the normalized flux (\spef)}
\tablefoottext{b}{For systems with multiple components, $z_{\rm \, DLA}$ is the weighted average of the absorption redshifts.} 
\tablefoottext{c}{Spectrum too noisy to identify metal absorption lines.} }
\end{landscape}
\end{longtab}

%
%________________________________________________________________

% for the bibliography, at the end
\bibliographystyle{aa} % style aa.bst
\bibliography{nlr_biblio} % your references Yourfile.bib

\end{document}